\documentclass[twocolumn]{aastex631}

\submitjournal{Astronomy Reports (Accepted)}

\usepackage{rotating} \usepackage{epsfig}
\usepackage{amsmath}
\usepackage{amssymb}
\usepackage{multirow}
\usepackage{graphicx}
\usepackage{float}
\usepackage{color}
\begin{document}

\title{THE OBSERVED POLARIZATION DIRECTION\\ DEPENDING ON GEOMETRICAL AND KINEMATIC PARAMETERS OF RELATIVISTIC JETS}

\author[0000-0001-7307-2193]{Marina S. Butuzova}
\affiliation{Crimean Astrophysical Observatory, Nauchny 298409, Russia}
\affiliation{Astro Space Center of Lebedev Physical Institute, Profsoyuznaya 84/32, Moscow 117997, Russia}

\begin{abstract}
The study of the polarization direction is crucial in the issue of restoring the spatial structure of the magnetic field in the active galaxy parsec-scale jets. But, due to relativistic effects, the magnetic field projected onto the celestial sphere in the source reference frame cannot be assumed to be orthogonal to the observed direction of the electric vector in the wave.
Moreover, the local axis of the jet component may not coincide with its motion direction, which affects the observed polarization direction. In this article, we analyze the transverse to jet distributions of the electric vector in the wave, obtained as a result of modeling with different jet kinematic and geometrical parameters for a helical magnetic field with a different twist angle and for a toroidal magnetic field in the center, surrounded by a varying thickness sheath, penetrated by a poloidal field.
We obtained: 1) the shape of the electric vector transverse distribution depends in a complex way on the angles of the jet axis and the velocity vector with the line of sight; 2) ambiguity in determining the twist direction of the helical magnetic field under using only the distributions of the electric vector in the wave; 3) both considered magnetic field topologies can reproduce both the ``spine-sheath'' polarization structure and individual bright details with the longitudinal to the jet axis polarization direction.
\end{abstract}


\section{Introduction} 

The phenomenon of active galactic nuclei (AGNs) is determined by the matter's accretion onto a supermassive black hole. This process causes the occurrence of an azimuthal component even with an initially poloidal magnetic field \citep[see, for example, numerical simulations][]{TchekhovskoyBromberg16}. The azimuthal component plays an important role in jet formation \citep{BlandfordZnajek77, BlandfordPayne82}. Therefore, it is natural to expect a presence of the ordered helical magnetic field at least up to distances of several parsecs from the true jet base. We can make some conclusions about the magnetic field  based on polarization observations. In particular, observations in the optical range detect sharp changes in the direction of the electric vector (EV) in the wave on different time scales without a well-established correlation with the total and polarized intensity and polarization degree \citep[][and references therein]{Raiteri19, Raiteri21}. On the one hand, the propagation of shocks through turbulence cells with a randomly oriented magnetic field can explain the observed variability of polarization properties \citep{Marscher14}. However, \citet{LyutikovKrav17} showed that under a strictly ordered helical magnetic field, a change in the jet speed and orientation relative to the line of sight leads to unsystematic changes in the polarization properties. 

Parsec-scale jets are directly observed by very long baseline interferometry (VLBI). In the data of these observations, there are changes in the polarization properties along and across jet \citep{ListerHoman05, Pushkarev17}. Recent results of the long-term monitoring of the parsec-scale AGN jets \citep{Pushkarev17, Pushkarev22}, carried out within the framework of the MOJAVE project (https://www.cv.nrao.edu/MOJAVE/index.html), indicate the presence of a stable distribution of polarization properties in the jets.
The observed transverse to jet distributions of total and polarized intensity, polarization degree, and EV deviation from the local jet axis are well reproduced in numerical simulations with a strictly ordered magnetic field and taking into account both the non-radial motion of components and the curved jet shape \citep{BP22PoS,BP22}. Moreover, there is evidence of an ordered helical magnetic field on kiloparsec scales \citep[for example, ][]{ Christodoulou16, Knuettel17}.

In addition to the helical magnetic field in AGN jets on parsec scales, it is often considered the ``spine-sheath'' topology. Namely, it is assumed the toroidal magnetic field is in the central part of the jet (spine), whereas the longitudinal one is in the external sheath surrounding the spine. Such magnetic field topology is confirmed by the observed distribution of EVs for several sources \citep{PushGab05ARep} and could arise due to the interaction of the jet flow with the environment \citep{Laing1980, Ghisellini05}. In recent simulations of the transverse to jet distributions of polarization properties for some parameters, a good correspondence has been found between the theoretical and observed profiles for both total and polarized intensity, polarization degree, and deviation of EV in the wave from the local jet axis for both the helical magnetic field and the ``spine-sheath'' configuration \citep{BP22PoS,BP22}.  

From the abovementioned it follows that the questions of whether the global magnetic field is present in the parsec-scale AGN jets, how ordered it is, and what its configuration is, remain open. The purpose of this article is, firstly, to identify in relativistic jets the main features of the helical magnetic field, which cannot be interpreted under the assumption of the ``spine-sheath'' topology, basing on the analysis of theoretical transverse distributions of the EV direction relative to the local jet axis, obtained in the investigations \citep{BP22PoS,BP22}. Secondly, to trace the dependence of the shapes of EV transverse distributions on the jet geometrical parameters. In Section~2, we give a brief description of the used geometrical and kinematic jet model, which most fully describes the observed properties \citep{Lister13, Lister21, Pushkarev17} of the considered objects. The analysis of the EV transverse distribution shapes and the identified main features for the helical field and the ``spine-sheath'' topology are given in Section~3. Sections~4 and 5 include a discussion of the results and conclusions, respectively.

\section{Jet model}

The results of long-term monitoring of the parsec-scale AGN jets carried out within the framework of the MOJAVE project reveal that the jet parts are injected at various position angles and often move along curved trajectories \citep{Lister13, Lister21}. At the same time, the jets of each object propagate within a strictly fixed angle on the celestial sphere \citep{PushkarevKLS17}. These data indicate that the conception of a straight jet, the velocity vector of which coincides with the jet axis, is outdated. Instead, a model of a helical jet with a non-radial motion of its components was proposed, which allowed us to describe the sometimes contradictory observed properties of blazars S5~0716+714 \citep{But18a, But18b, But21Ap} and OJ~287 from parsecs to kiloparsecs scales \citep{BP20, But21ARep}. This model, with a wide variation of parameters, was used to calculate transverse distributions of polarization properties, some of which correspond well to the observed ones \citep{BP22PoS,BP22}.
It is important to emphasize that the simulation was carried out under the assumption of an optically thin jet for the following reasons. First, it was shown that the optical thickness of $\tau>1$ only in VLBI cores (the most compact and bright feature on the radio map), with the distance from which $\tau$ decreases sharply \citep{Pushkarev12}. Second, the Faraday rotation for jets from the MOJAVE sample is only a few degrees and mainly occurs in the region close to the VLBI core \citep{Hovatta12}, which indicates a low number density of thermal electrons, the influence of which on the polarization properties observed at 15~GHz can be neglected.

For the possibility of analytical description, the orientation changes of the jet local axis in space, for simulation we assumed a helical jet, the axis of which lies on the surface of an imaginary cone with a half-solution angle $\xi=1^\circ$. The jet was divided into cylindrical components, which axis and velocity vector make the angle $\rho$ and $p$ with the cone generatrix, respectively. The azimuthal angle of the consecutive components $\varphi$ is different.
The $\varphi$ change leads to changes in the angle $\theta$ between the component velocity vector and the line of sight, reflected in the value of the Doppler factor $\delta$ and the angle at which the jet component is located relative to the observer $\theta_\rho$. 
By assigning different values to the model parameters, namely, the angle of the cone axis with the line of sight $\theta_0=2$, 5, and 10$^\circ$, $p=2$, 3, 5, and 10$^\circ$, and $\rho/p=1$, 2, 3, 5, 15, and 25, under the condition $\rho<90^\circ$, we obtained 63 sets of geometrical parameters of the jet.
For each of them, the magnetic field was set to be helical with a twist angle $\psi^\prime=0$, 10, 25, 45, 55, 75, and 90$^\circ$ and the ``spine-sheath'' topology with a distance from the axis at which the transition occurs equal to $R_t=0.25$, 0.33, 0.5, 0.7 and 0.9 in jet radius units.
The speed of the components was assumed to be 0.995$c$ ($c$ is the speed of light). Thus, we considered a total of 819 model parameter sets. For each of them, taking into account the relativistic effects \citep{LPG05}, the Stokes parameters were calculated by integrating along the line of sight at 61 equidistant points, which, when projected onto the picture plane, are located on the cross-section of the jet component and are in the range from $-0.9$ to 0.9 jet radii. Then, the resulting distribution was convoluted with a one-dimensional Gaussian with a full width at half maximum equal to 1/3 of the component width, and transverse distributions of total and polarized intensity, polarization degree, and deviation of EV in the wave from the local jet axis were constructed. It is important to note the  falseness of restoring the magnetic field in the source as perpendicular to the observed EV, since this method does not account relativistic effects \citep{LPG05}.Therefore, in this article we analyze the shape of the obtained transverse distributions of the EV direction depending on the geometrical parameters for given configurations of the magnetic field in the jet.

\section{Analysis of transverse distribution of polarization direction}
In the observational data \citep{ListerHoman05, PushGab05ARep}, main types of transverse distributions of EVs are distinguished. Namely, 1) EV is directed parallel to the jet axis; 2) EV is perpendicular to the axis; 3) EV is parallel on one side of the jet and perpendicular on the other; 4) EV is parallel near the jet axis and perpendicular at the edges. In the latter case, the polarization structure is called a ``spine-sheath'', but we do not directly associate it with the ``spine-sheath'' topology of the magnetic field. The main types of EV distributions considered by us and the criteria for their determination based on the data of simulations performed by \citet{BP22PoS,BP22} are given in Table~1.

\begin{table*}
\caption{The main types of transverse EV distribution and criteria for their determination.}
\bigskip
\begin{tabular}{|c|c|l|}
\hline
 Number & Brief description & Definition criteria \\
\hline
1 & transverse & The average of the points in cut $>60^\circ$ and \\
 ~ & ~ & the amplitude of the changes $\left(\Delta \text{EV}\right)\leqslant 20^\circ$ \\
 \hline
2 & longitudinal & The average of the points in cut $<30^\circ$ and \\
~ & ~ & $\Delta \text{EV}\leqslant 20^\circ$ \\ 
\hline
3 & ``oblique'' & $30^\circ \leqslant$ average for all points $\leqslant 60^\circ$ \\
\hline
4 & ``spine-sheath'' & Average on segments [$-0.9$, $-0.3$] and [0.3,0.9] > by 30$^\circ$\\
~ & ~ &average on segment  [$-0.39$,0.39] and \\
~ & ~ & minimum on the segment  [$-0.39,0.39$]  $< 45^\circ$\\
\hline
5 & ``left-asymmetric'' & Average on segment [$-0.9$, $-0.3$) < average on segment\\ 
~ & ~ &  [$-0.3$, 0.3) < average on segment [0.3,0.9] and \\
~ & ~ & $\Delta \text{EV}>30^\circ$ \\
\hline
6 & ``right-asymmetric'' & Average on segment [$-0.9$ ,$-0.3$) > average on segment \\
~&~& [$-0.3$, 0.3) > average on segment [0.3,0.9] and \\
~&~& $\Delta \text{EV}>30^\circ$\\
\hline
7 & undefined & Other ones that do not correspond to types 1-6\\
\hline
\end{tabular}
\end{table*}

The shape of the observed transverse distribution of EVs depends on $\delta$ and the angle between the component axis and the line of sight $\theta_\rho$. Since we assume the component speed is unchanged, the dependence on $\delta$ is reduced to that on $\theta$. For $p\neq 0^\circ$, the range of $\theta$ values varies in a larger range than $\theta_0\pm\xi$ \citep{But18a}.
This interval depends on $\theta_0$ and $p$ and is individual for each parameter set. To display the change in the shape of the EV distribution depending on the angle, for each considered parameter set (or sets) we divided the range of $\theta$ values into three parts (small, medium, and large values of $\theta$ angles), for each of which we plotted a histogram of the occurrence of the selected distribution types (see Appendix, Fig.~\ref{fig:fig5}-\ref{fig:fig17}). To eliminate the influence of the uneven distribution of $\theta$ values in the considered interval, we divided the number of EV distributions of a certain type by their total number for each $\theta$ interval. Solid, dashed and dotted lines correspond, in the order of enumeration, to the large, medium and small $\theta$. Analyzing the obtained histograms, the following regularities can be identified.

\subsection{Helical magnetic field}
For $\psi^\prime=0^\circ$,  the EV direction transverse to the jet significantly prevails (Appendix, Fig.~\ref{fig:fig5}). With increasing both $\rho/p$ and $p$, longitudinal and oblique to the jet axis distributions of EV directions appear firstly for small, then for large angles (Appendix, Fig.~\ref{fig:fig5}). The situation changes violently already for $\psi^\prime=10^\circ$ (Appendix, Fig.~\ref{fig:fig6}).
Transverse EVs are mainly found at $\rho/p=1$, whereas the dominant type of EV distribution becomes ``right-asymmetrical''. For $\rho/p\geqslant15$ and $p\geqslant5^\circ$, ``left-asymmetrical'' EV distributions also begin to occur at all intervals of $\theta$ values.
Moreover, only for $p=10^\circ$ and $\rho/p\geqslant5$, at large $\theta$, a significant number of the ``spine-sheath'' EV distribution type appear. A similar situation is observed for $\psi^\prime=25^\circ$ (Appendix, Fig.~\ref{fig:fig7}). 
It is interesting to note that, for example, at $p=10^\circ$ and$\rho/p\geqslant3$, the ``left-asymmetrical'' type of EV distribution is present for large $\theta$ values, and ``right-asymmetrical'', one for medium and small values, whereas at $p=3^\circ$ and $\rho/p=25$ ``right-'' and ``left-asymmetrical'' EV distributions occur at all values of $\theta$ angles.
That is, for the same magnetic field configuration and the same angle of the jet component velocity vector with the line of sight, the observer can register both ``right-'' and ``left-asymmetrical'' EV distributions. The differences in distributions may be caused by different angles between the jet component axis and the line of sight ($\theta_\rho$) at the same $\theta$. 

\begin{figure*}
\includegraphics[scale=0.9]{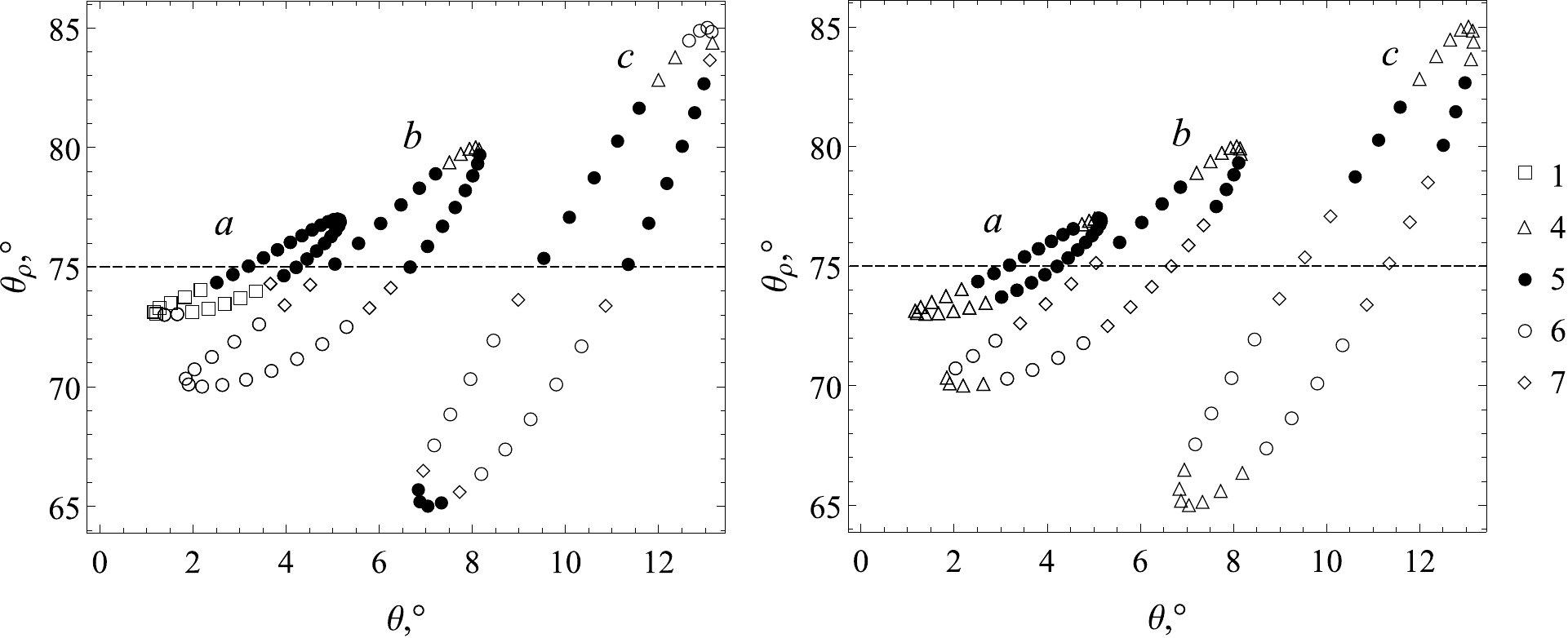} 
\caption{
Types of EV distribution (marked with different symbols) depending on $\theta$ and $\theta_\rho$ for $p=3^\circ$ and $\rho/p=25$. The points forming closed loops \textit{a}, \textit{b} and \textit{c} are obtained for $\theta_0=2$, 5 and 10$^\circ$, respectively, and equidistant values of the azimuth angle. The dotted line is drawn for $\theta_\rho=p$. The angle of the helical magnetic field with the local jet axis is 25$^\circ$ (left) and 65$^\circ$ (right).}
\label{fig:fig1}
\end{figure*}

To test this assumption, we plotted the occurrence of types of the EV distribution shapes as a function of $\theta$ and $\theta_\rho$ (Fig.~\ref{fig:fig1}). The angle $\theta_\rho$ was calculated according to the formulas (11)-(13) in \citep{But18a}, substituting $\rho$ instead of $p$.
Due to the discreteness of the parameter values in our simulation, three closed loops \textit{a}, \textit{b}, and \textit{c} appear on the left panel of Fig.~\ref{fig:fig1}, whose points correspond to 36 equidistant values of the azimuthal angle $\varphi$ for $\theta_0=2$, 5, and 10$^\circ$, respectively. For loop \textit{a} ``right-asymmetrical'' EV distributions (type~6) occur only at the minimum possible values of $\theta_\rho$, while for $\theta_\rho\gtrsim 74^\circ$ there is only the ``left-asymmetrical'' EV distribution (type~5) is present. 
Similar behavior is observed in loops \textit{b} and \textit{c}, except for two facts. First, the frequency of occurrence of ``right-asymmetrical'' EV distributions increases. Second, for the loop \textit{c} with the maximum amplitude of variation $\theta_\rho$ for $\theta_\rho<75^\circ$, in addition to the ``right-asymmetrical'' type, there is also the ''left-asymmetrical'' type of EV distribution at the minimum achievable $\theta_\rho$ angles.
On the other hand, for $\theta_\rho>75^\circ$, the ``left-asymmetrical'' EV distribution prevails, but at the maximum attainable angles $\theta_\rho$, the ``right-asymmetrical'' one occurs. It is important to note that for the same Doppler factor but different $\theta_\rho$, different types of EV distributions will be registered. 
For example, for $\theta=3^\circ$, with increasing $\theta_\rho$ from 69$^\circ$ to 78$^\circ$, the ``right-asymmetrical'', longitudinal, and ``left-asymmetrical'' EV distributions are observed for the same helical magnetic field with the twist angle $\psi^\prime=25^\circ$ in the source. Thus, the complex dependence of the EV distribution shape on $\theta$ and $\theta_\rho$ makes it impossible to define of twisting direction of the jet helical magnetic field based only on the EV distribution asymmetry.

With a further increase of $\psi^\prime$ up to 45$^\circ$, the transverse EVs at $\rho/p>1$ almost entirely disappear, the ``left-asymmetrical'' EV distribution begins to prevail and the occurrence of the ``spine-sheath'' structure increases (Appendix, Fig.~\ref{fig:fig8}). The tendency to the dominance of the number of distribution cases with the 4th and 5th types is present up to $\psi^\prime=65^\circ$ (Appendix, Fig.~\ref{fig:fig10}), under which the prevail EV distribution is the ``spine-sheath'' structure, presenting in almost all ranges of values of $\theta$ and $\theta_\rho$  (Fig.~\ref{fig:fig1}, right panel). When the magnetic field approaches the toroidal one ($\psi^\prime=90^\circ$), the EV distribution of the ``spine-sheath'' type is present in the majority of cases (Appendix, Fig.~\ref{fig:fig12}).

\begin{figure*}  
\includegraphics{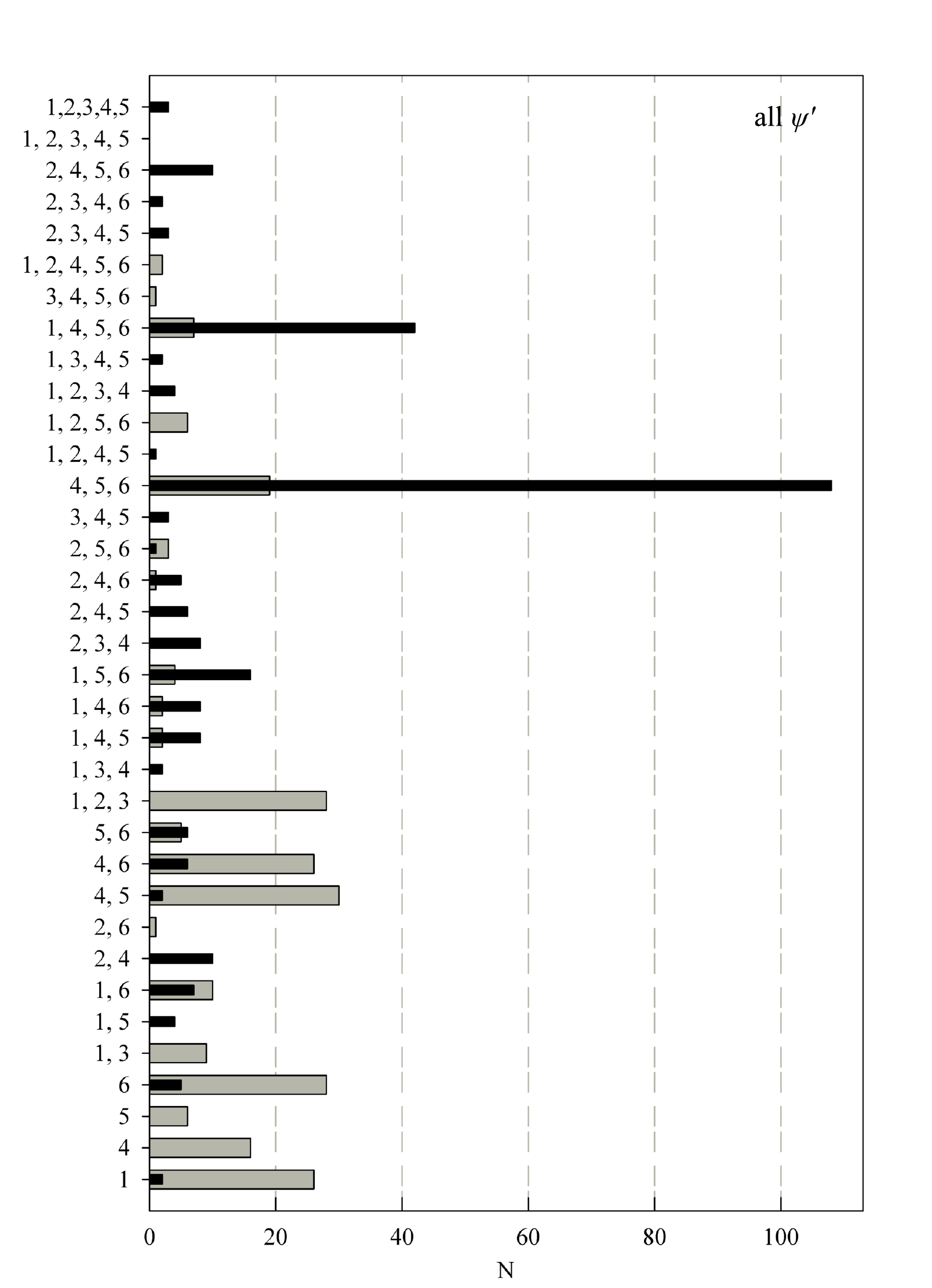}
\caption{Histogram of the occurrence of EV distribution shape combinations in each model jet for all values of $\psi^\prime$. Combinations of types of EV distribution shapes without and additionally containing type 7 are marked in gray and black, respectively.}
\label{fig:fig2}
\end{figure*}

Figure~\ref{fig:fig1} shows that the fixed sets of model parameters can reproduce several types of EV distribution shapes. We plotted histograms (Appendix, Fig.~\ref{fig:fig18}-\ref{fig:fig25}) to analyze the number of EV distribution shapes and their types presented in an individual model jet. For avoid the figure bulkiness, the same combinations of EV distribution shapes without and with type~7 were plotted together. Figure~\ref{fig:fig2} shows a histogram for all considered $\psi^\prime$ values. It is seen that about half of the cases have only 1 or 2 EV distribution shapes. 
About one-third of all cases have 4 or more shapes of EV distributions. In the majority of model jets there are transverse, ``right-'' and ``left-asymmetrical distributions, the ``spine-sheath''structure, and their combinations. It is necessary to emphasize that types~5 and 6 in the considered right-hand helical magnetic field are equally occurring; moreover, under several sets of model parameters, these shapes of EV distributions are present in one jet. This fact indicates the ambiguity of the definition of the magnetic field direction based only on the EV distributions.

Additionally, we note that regions with the longitudinal direction of EVs (type~3), interpreted only under the assumption of the shock wave in the jet transverse to the jet axis \citep[see, e.g.,][]{Laing1980, Hughes89}, arise in some model jets and always in combination with other distribution types (see Fig.~\ref{fig:fig2}). Figures~\ref{fig:fig5}-\ref{fig:fig17} in Appendix show that type~3 occurs only under$\rho/p>1$ and $\psi^\prime=0$ or 90$^\circ$. For a longitudinal field, type~3 generally accounts for all intervals of values of the velocity vector angle with the line of sight, although with some parameters, only for small or medium  $\theta$ values. Whereas for the toroidal field, type~3 occurs only at small and medium angles, which implies a larger Doppler factor, and, as a consequence, a high radiation intensity in these regions with longitudinal EVs.

\subsection{Spine-sheath topology}

Under a small ``spine'' radius of  $R_t=0.25$ (in units of the jet radius) containing the toroidal magnetic field, the penetrated by the longitudinal magnetic field ``sheath'' mainly influences the character of transverse EV distribution (Appendix, Fig.~\ref{fig:fig13}). It leads to the dominance of transverse EVs and a small difference from the case of $\psi^\prime=0^\circ$. For $\rho/p\geqslant3$, a certain number of other types of EV distributions appear, the occurrence of which, especially of the ``spine-sheath'' type, increases with increasing $R_t$. For $R_t=0.5$ (Fig.~\ref{fig:fig15}), the ``spine-sheath'' distribution shape becomes dominant. With a further increase in $R_t$ (Appendix, Fig.~\ref{fig:fig17}), the dominance of the ``spine-sheath'' type increases, but transverse EVs are mainly present at $\rho/p=1$.

\begin{figure*}
\includegraphics[scale=0.8]{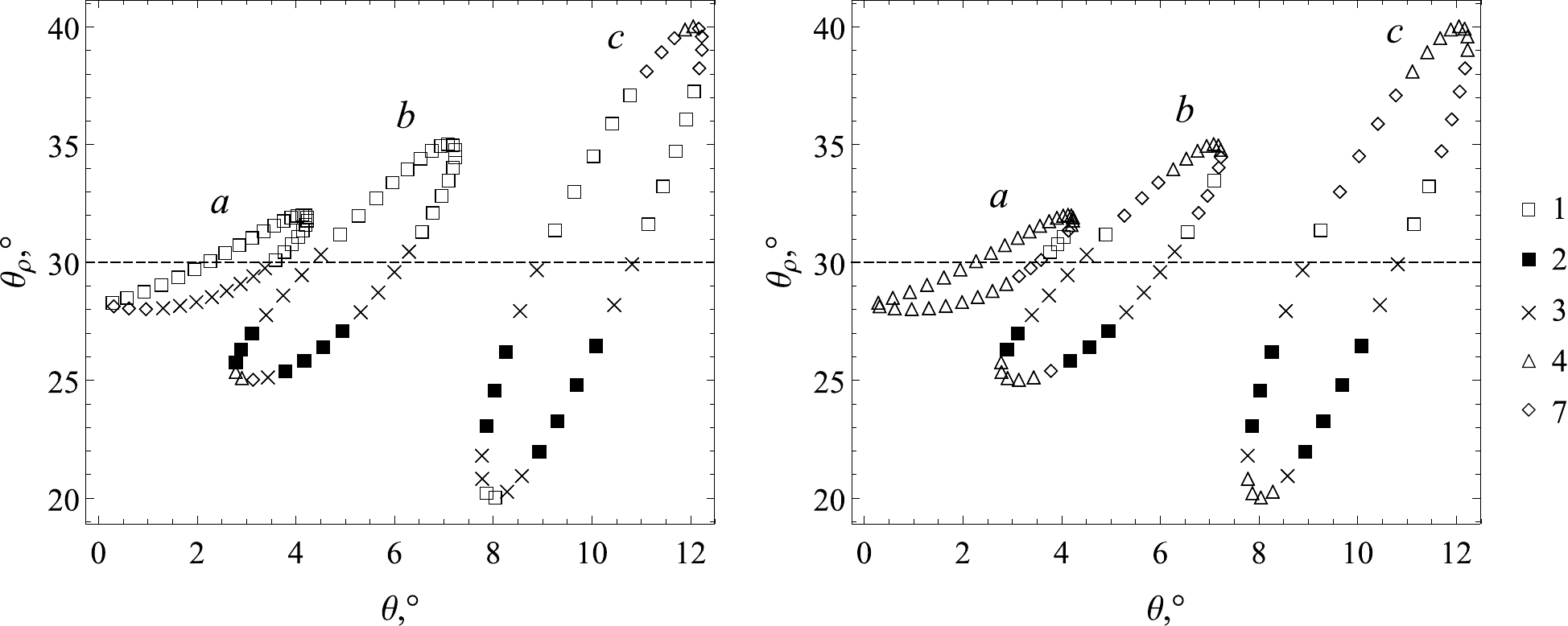} 
\caption{
The dependence of the EV distribution type on $\theta$ and $\theta_\rho$ for $p=2^\circ$, $p/\rho=15$ and $R_t=0.33$ (left) and 0.5 (right). The points forming closed loops \textit{a}, \textit{b} and \textit{c} were obtained for 36 equidistant values of the azimuthal angle $\varphi$ for $\theta_0=2$, 5, and 10$^\circ$, respectively. The dotted line marks $\theta_\rho=p$.}
\label{fig:fig3}
\end{figure*}

Figure~\ref{fig:fig3} shows that for $R_t=0.33$ and $\theta_\rho\gtrsim 30^\circ$, longitudinal EVs prevail, whereas, with an increase in the ``spine'' width and, consequently, with an increase in its contribution to the total radiation, the ``spine-sheath'' structure manifests. On the other hand, for both considered $R_t$ at $22^\circ\lesssim \theta_\rho \lesssim 27^\circ$, there are longitudinal EVs. However, as the same case of the helical magnetic field, there is no explicit dependence of the shape of the EV distribution, since, for example, the transverse EVs is also found for the minimally achievable $\theta$ and $\theta_\rho$ at $R_t=0.33$ and $\theta_0=10^\circ$. In the case of $R_t=0.5$ and $\theta_0=10^\circ$, the ``spine-sheath'' distribution occurs for both minimum and maximum values of the angles $\theta$ and $\theta_\rho$.

Considering the number of different types of EV distributions in the model jet, we can say the following. Approximately 2/3 of the parameter sets reproduce no more than 3 types of EVs distributions, but model jets with only one type prevail (Fig.~ref{fig:fig4}). Namely, it is either the ``spine-sheath'' polarization structure or the transverse EVs, occurring for 20\% and 10\% of cases, respectively. Noteworthy, it is rarely, but for the symmetric magnetic field in the source reference frame, ``right-'' or ``left-asymmetrical'' EV distributions appear, but only in combination with other types under the fixed model parameter set.
Longitudinal EVs occur only together with other types at $\rho/p>1$ and $R_t\leq0.5$ for the entire range of $\theta$, but with a tendency to increase in occurrence at small $\theta$. Combinations of EV distribution shapes at different ``sheath'' thicknesses are presented in Appendix, Fig.~\ref{fig:fig26}-\ref{fig:fig30}. 

\begin{figure*} 
\includegraphics{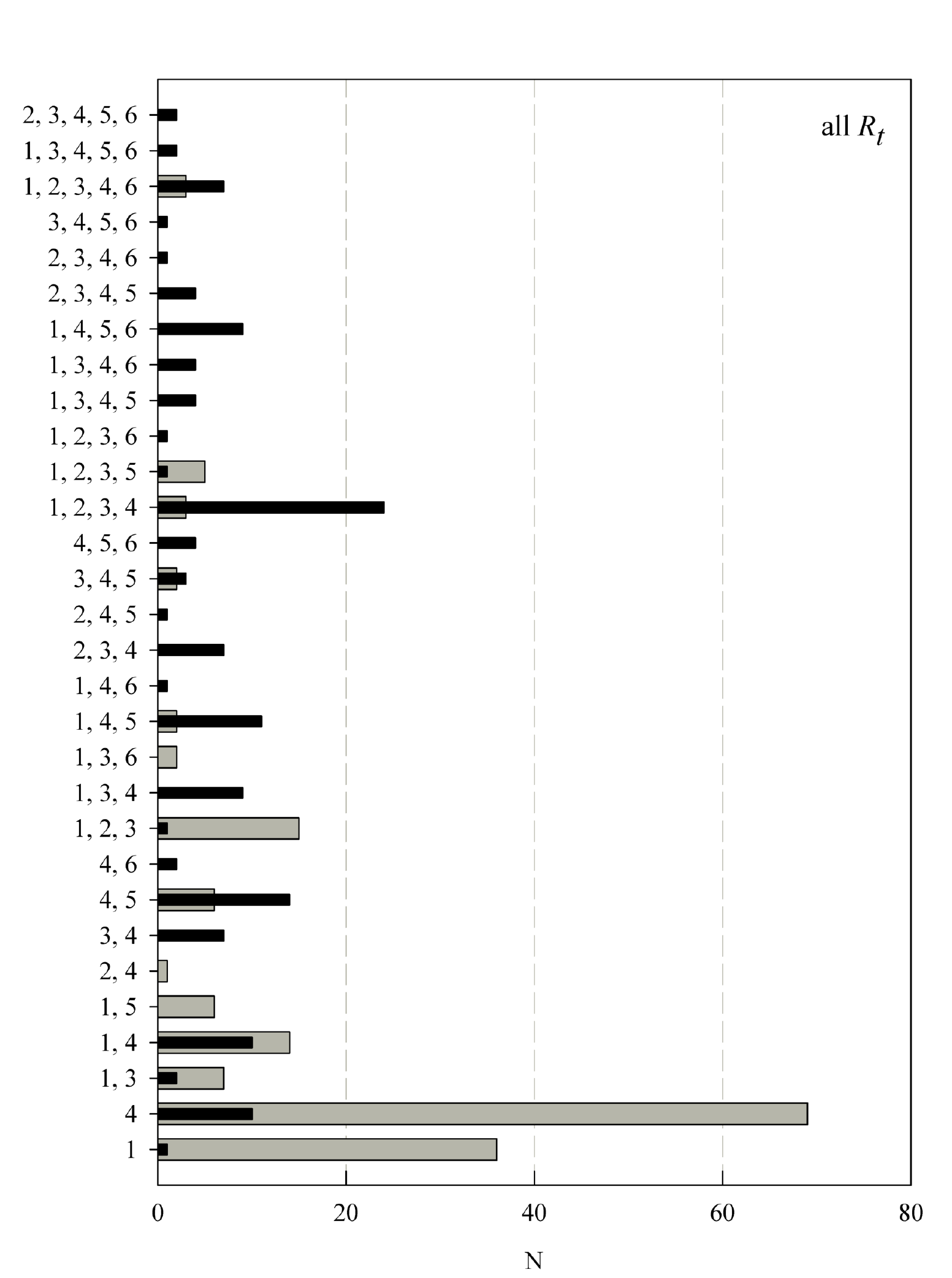}
\caption{Combinations of EV distribution shapes in each model jet for all $R_t$ values. Those that additionally contain type 7 are marked in black.}
\end{figure*}

\section{Discussion}

The definition of the magnetic field configuration in AG jets is based on the measurement of the Faraday rotation of the polarization plane and on the analysis of the EV distribution in the jet. The results of the first method can only be applied indirectly to a jet, since a noticeable change in the polarization direction occurs only when a wave propagates in a thermal plasma, probably surrounding a relativistic jet \citep{Gabuzda14}. For instance, \citet{Gabuzda15, Gabuzda18} found a significant transverse jet gradient of the rotation measure in most studied objects. Moreover, for some of them, the Faraday rotation measure has a different sign on different jet sides, which directly indicates the opposite directions of the magnetic field on different jet sides. These facts allowed the authors to conclude about the presence of a helical magnetic field in the jet itself. This conclusion was further confirmed by the fact that objects with the detected transverse gradient of the rotation measure have EVs longitudinal at the jet axis and transverse at one or two sides \citep{Gabuzda14}.
The origin of the ``spine-sheath'' polarization structure is explained by the helical field directed at a large angle to the jet axis, while a field with a small twist angle creates asymmetric distributions. The latter is due to the influence of relativistic effects \citep{LPG05} and geometry, which give reason to expect longitudinal EVs on the side of the jet where the magnetic field is directed closer to the line of sight and transverse EVs on the opposite side \citep{Gabuzda21}.
Therefore, the magnetic field direction in the jet can be defined by the asymmetry of the EV distributions. However, all the analysis of observational data and modeling until recently were carried out under the assumption that the jet axis coincides with the velocity vector \citep[e.g.,][]{LPG05, MurphyCaw13, PriorGourg19}. Therefore, the observed change in the transverse EV distribution was associated with the change in either the jet angle with the line of sight or the twist angle of the magnetic field.

For the first time in simulations, \citet{BP22PoS,BP22} considered the different values of the angles of the jet velocity vector and the local axis with the line of sight. Within this framework, at $\rho/p\neq1$, it was obtained the Stokes parameter $U\neq0$, which provides EVs deviations from the local jet axis that differ from 0$^\circ$ and 90$^\circ$, which expected by \citet{LPG05}. 
The discrepancy of the observed EVs with longitudinal or transverse directions was interpreted either by introducing additional assumptions, for example, about the disorder of the magnetic field \citep{MurphyCaw13}, or was decreased by accounting for the Faraday rotation \citep{HutchisonCaw01}. Since the objects from the MOJAVE sample are negligibly affected by the Faraday rotation \citep{Hovatta12}, we did not take this effect into account in the analysis.
We have shown that in most cases of simulation results, the observed typical transverse to jet distributions of EVs are reproduced. The presence of EV distributions that do not correspond to the typical ones (number~7 in Table~1) is associated with a strong asymmetry that occurs at $\psi^\prime \approx \left(45^\circ \pm 10^\circ \right)$ and makes it difficult to determine the shape of the EV distribution.
For the ``spine-sheath'' magnetic field topology, under some combination of parameters, it occurs a transverse EV distribution occurs, containing either perpendicular EVs on the axis and parallel at the edges or the described ``spine-sheath'' structure occupying 3/4 of the jet width, surrounded at the edges by areas with longitudinal EVs.
Taking into account that the simulations were performed under the assumption of the strictly ordered magnetic field and there is a correspondence with the observed transverse distributions of total and polarized intensity, polarization degree, and EVs for quasar jets 0333+321 (NRAO~140), 0836+710 (4C~+71.07), and 1611+343 \citep{BP22PoS,BP22}, we have convincing reasons to believe that the magnetic field on parsec-scales is highly ordered.

We analyzed the influence of the Doppler factor on the shape of the EV transverse distribution at the different magnetic field configurations. Since, in the simulation, the jet component velocity was unchanged and equal to $\beta=0.995$ in units of the speed of light (which corresponds to the Lorentz factor 10), we constructed distributions of the occurrence of relevant types of EV distributions depending on the angle between the velocity vector and the line of sight $\theta$.
Note that at ultrarelativistic velocities, the effect of $\theta$ on the Doppler factor is higher than $\beta$. Therefore, a change in $\beta$ would lead to a minor change in $\delta$, which would become the same as with $\beta=0.995$, but with a slightly different $\theta$.
The value of this angle for given parameters would most likely remain in the same part of the range of possible T values, into which we divided to plot histograms of the occurrence of relevant types of EV distribution. Even if in some cases, the new value of $\theta$ would correspond to another part of the interval, it would not qualitatively change the results obtained here.

\section{Conclusions}

We analyzed the transverse to the jet profiles of EV distributions in the observer's reference frame under the given configuration of the global magnetic field and various kinematic and geometrical jet parameters. The main conclusions are as follows.

1) Changing the angles of the jet component axis and its velocity vector with the line of sight leads to a change in the observed profile of the transverse EV distribution under the constant magnetic field.

2) The characteristic types of EV distributions for the longitudinal magnetic field $\psi^\prime=0^\circ$) and the ``spine-sheath'' configuration are similar, the predominant type is the transverse EVs, but there are longitudinal and oblique ones.

3) Both considered magnetic field topologies reproduce the ``spine-sheath'' polarization structure, in which the EVs are parallel in the jet center, and perpendicular to the local jet axis at the edges. This distribution type, along with the `` left-asymmetric'' one, is dominant for the helical field with $\psi^\prime=45^\circ$, and starting $\psi^\prime=55^\circ$ and above, that is the main dominant type of distribution. For the ``spine-sheath'' filed topology at $R_t\leq0.33$, the polarization structure of the ``spine-sheath'' noticeably begins to manifest itself only for $p\geq 5^\circ$ and $\rho/p\geq3$, while transverse jet EVs remain dominant. At $R_t\geq0.5$, the ``spine-sheath'' EV distribution is significantly dominant for all other model parameters. Despite this, transverse EVs continue to occur at $R_t\geq0.5$ only for $\rho/p=1$.

4) The right-hand direction of the magnetic field twist with $\psi^\prime=10^\circ$ can give ``right-'' and ``left''-asymmetric'' stable observed EV distributions in a single source. At large $\psi^\prime$ up to 75$^\circ$, distributions of the 5th and 6th types are not observed simultaneously in model jets; they are present in combination with the ``spine-sheath'' structure or more complex ones. For some fixed model parameters for $\psi^\prime=10$ and 25$^\circ$, the sources can have only a ``right-'', or only a ``left-asymmetric'' EV distribution. All of the above indicates that to determine the direction of helical magnetic field twist, it is necessary to accounting the shapes of the transverse EV distributions along the entire jet, and the geometrical and kinematic parameters of the flow.

5) For both the helical magnetic field and ``spine-sheath'' topology, the longitudinal to jet axis EVs are reproduced in some model parameter sets and only in combination with other distribution types. Longitudinal EVs tend to occur at high Doppler factors, which leads to the detection of bright jet features with longitudinal EVs, which were previously associated only with shock waves.          

Thus, in order to obtain reliable conclusions about the physical conditions in jets based on radio-interferometric observations, the study of polarization properties must be supplemented with the results of kinematics and geometry analysis.

\section{Acknowledgments}
This work was supported by the Russian Science Foundation grant No. 21-12-00241.

\bibliography{Polarization_ARep}{}
\bibliographystyle{aasjournal}

\appendix

\section{EV distributions for different model parameters and combinations of EV distribution shapes in an individual jet}

\begin{sidewaysfigure}
\centerline{\epsfig{file=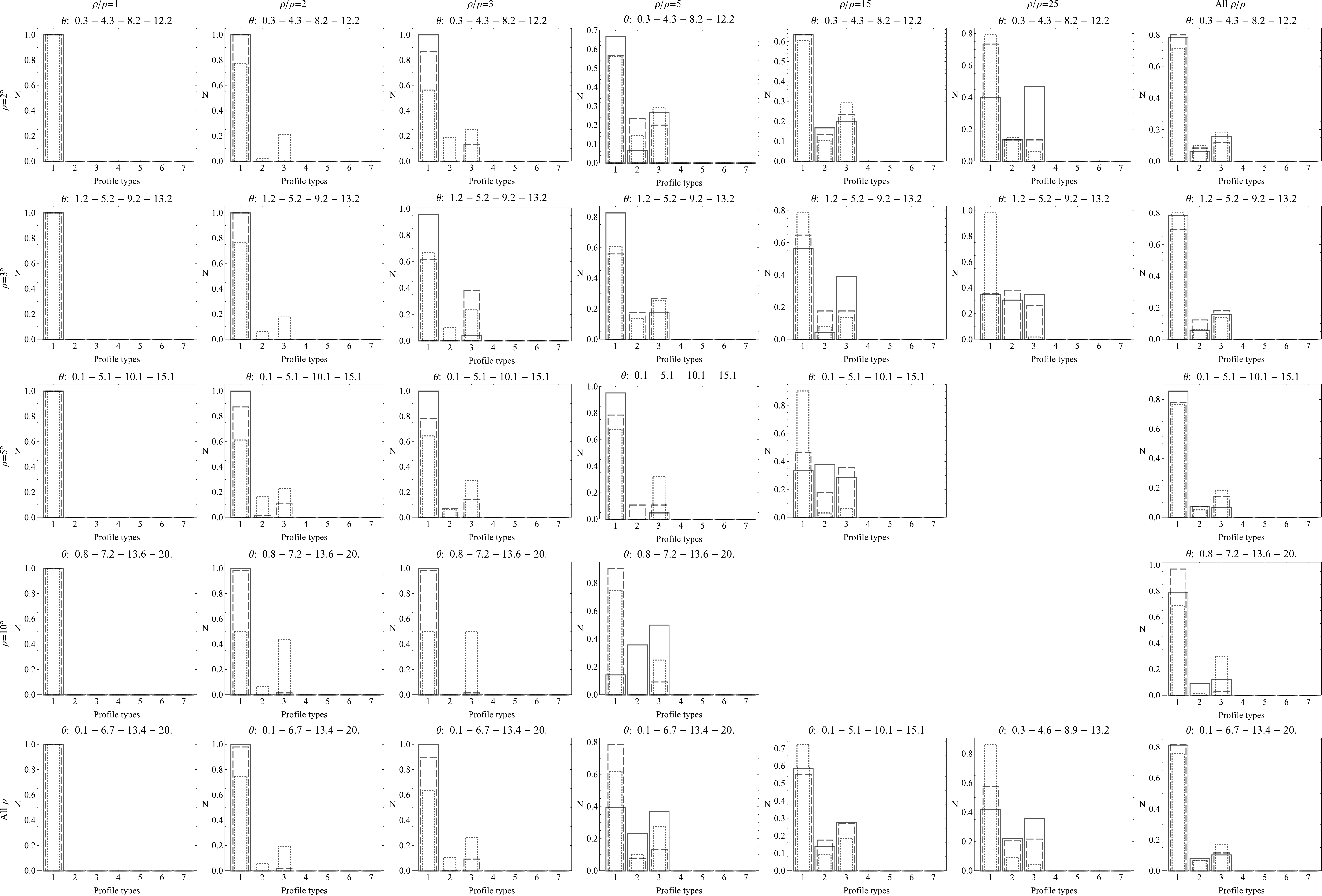, scale=0.35}}
\caption{Shapes of EV distributions depending on the angle of the velocity vector ($\theta$) and the jet component axis ($\theta_\rho$) to the line of sight for $\psi^\prime=0^\circ$. Solid, dashed, and dotted lines are associated with intervals of high, medium, and small values, respectively. The intervals of $\theta$ are indicated at the top of each plot.}
\label{fig:fig5}
\end{sidewaysfigure}

\begin{sidewaysfigure}
\centerline{\epsfig{file=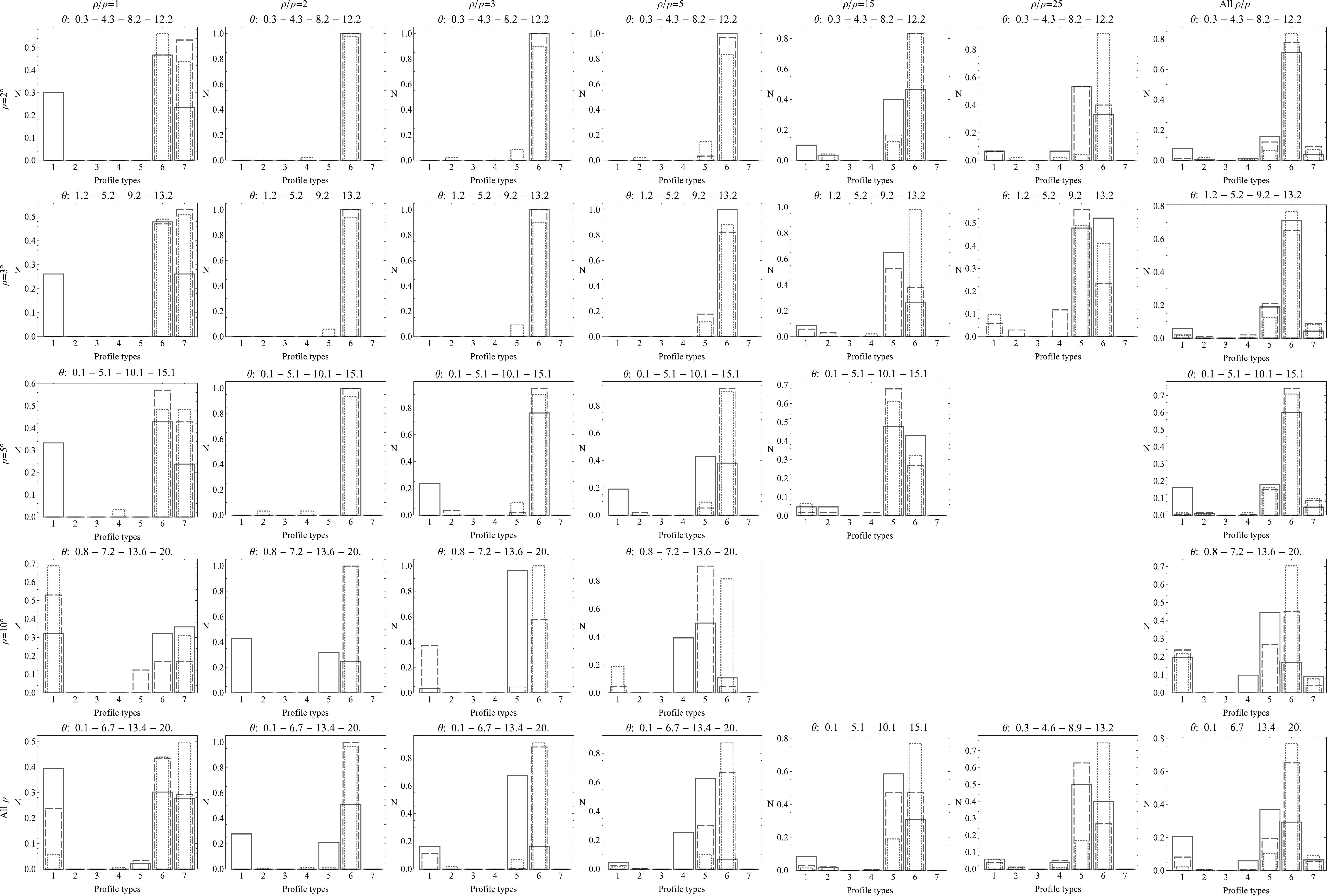, scale=0.35}}
\caption{Shapes of EV distributions depending on the angle of the velocity vector ($\theta$) and the jet component axis ($\theta_\rho$) to the line of sight for $\psi^\prime=10^\circ$. Solid, dashed, and dotted lines are associated with intervals of high, medium, and small values, respectively. The intervals of $\theta$ are indicated at the top of each plot.} 
\label{fig:fig6}
\end{sidewaysfigure}

\begin{sidewaysfigure}
\centerline{\epsfig{file=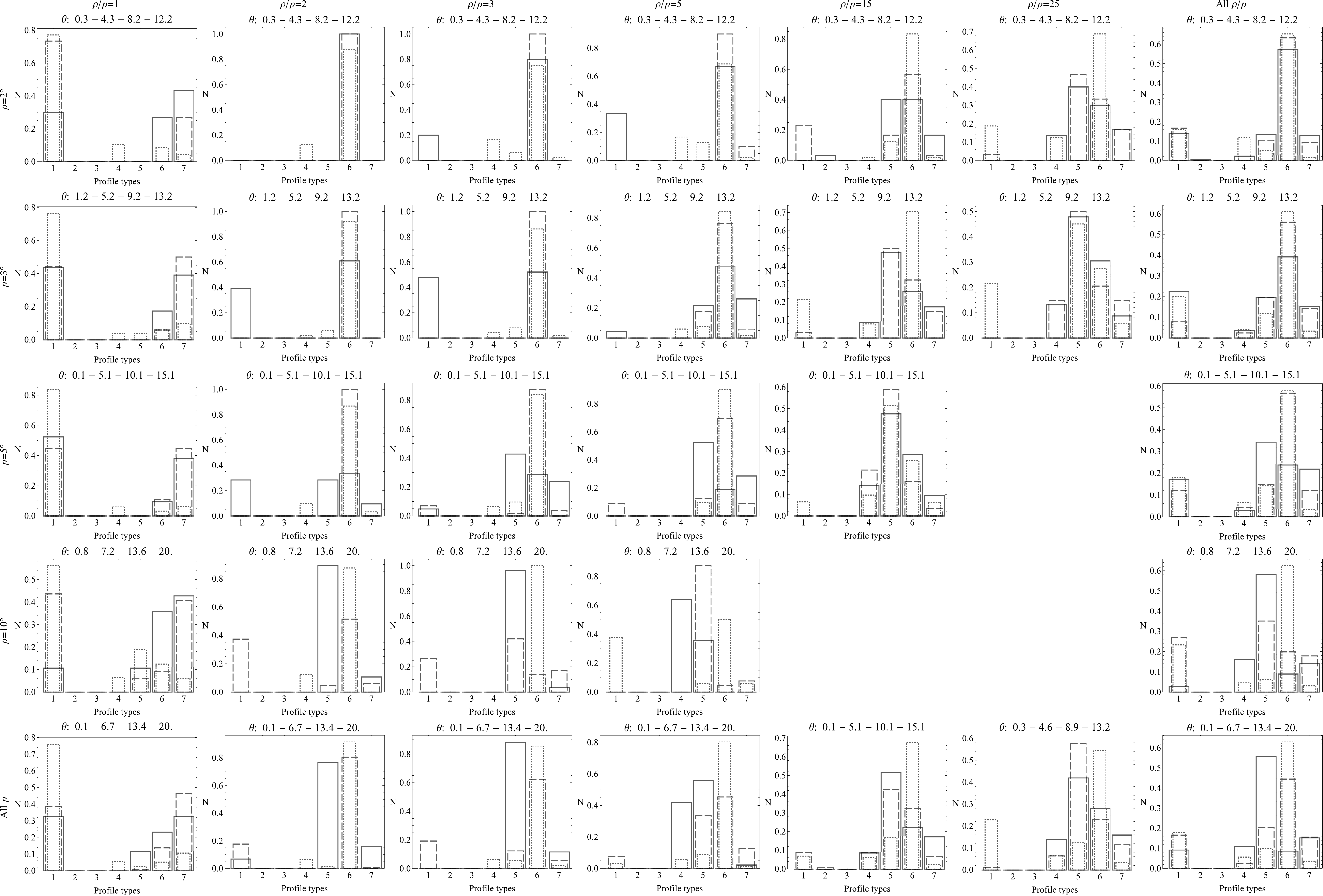, scale=0.35}}
\caption{Shapes of EV distributions depending on the angle of the velocity vector ($\theta$) and the jet component axis ($\theta_\rho$) to the line of sight for $\psi^\prime=25^\circ$. Solid, dashed, and dotted lines are associated with intervals of high, medium, and small values, respectively. The intervals of $\theta$ are indicated at the top of each plot.}
\label{fig:fig7}
\end{sidewaysfigure}

\begin{sidewaysfigure}
\centerline{\epsfig{file=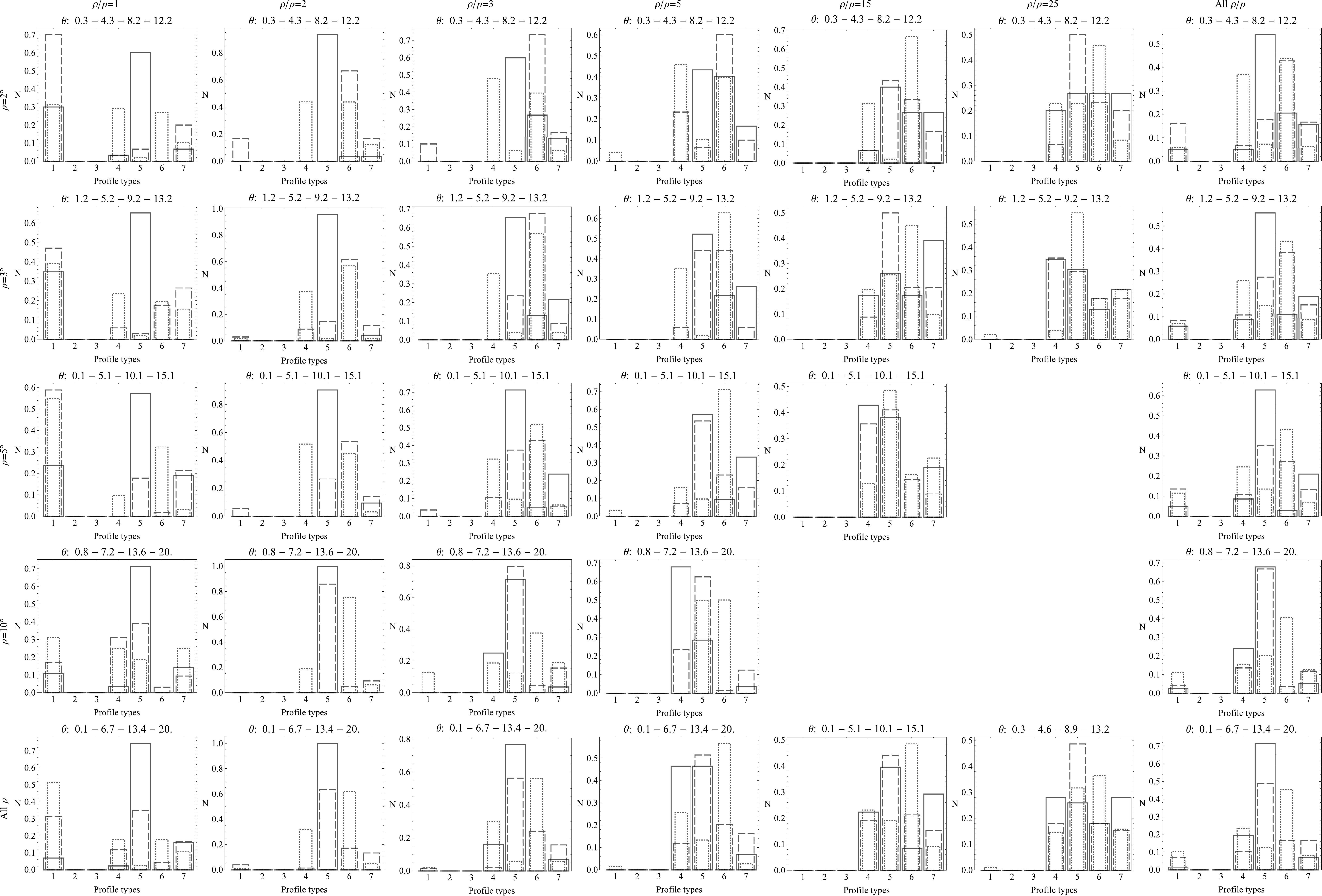, scale=0.35}}
\caption{Shapes of EV distributions depending on the angle of the velocity vector ($\theta$) and the jet component axis ($\theta_\rho$) to the line of sight for $\psi^\prime=45^\circ$. Solid, dashed, and dotted lines are associated with intervals of high, medium, and small values, respectively. The intervals of $\theta$ are indicated at the top of each plot.}
\label{fig:fig8}
\end{sidewaysfigure}

\begin{sidewaysfigure}
\centerline{\epsfig{file=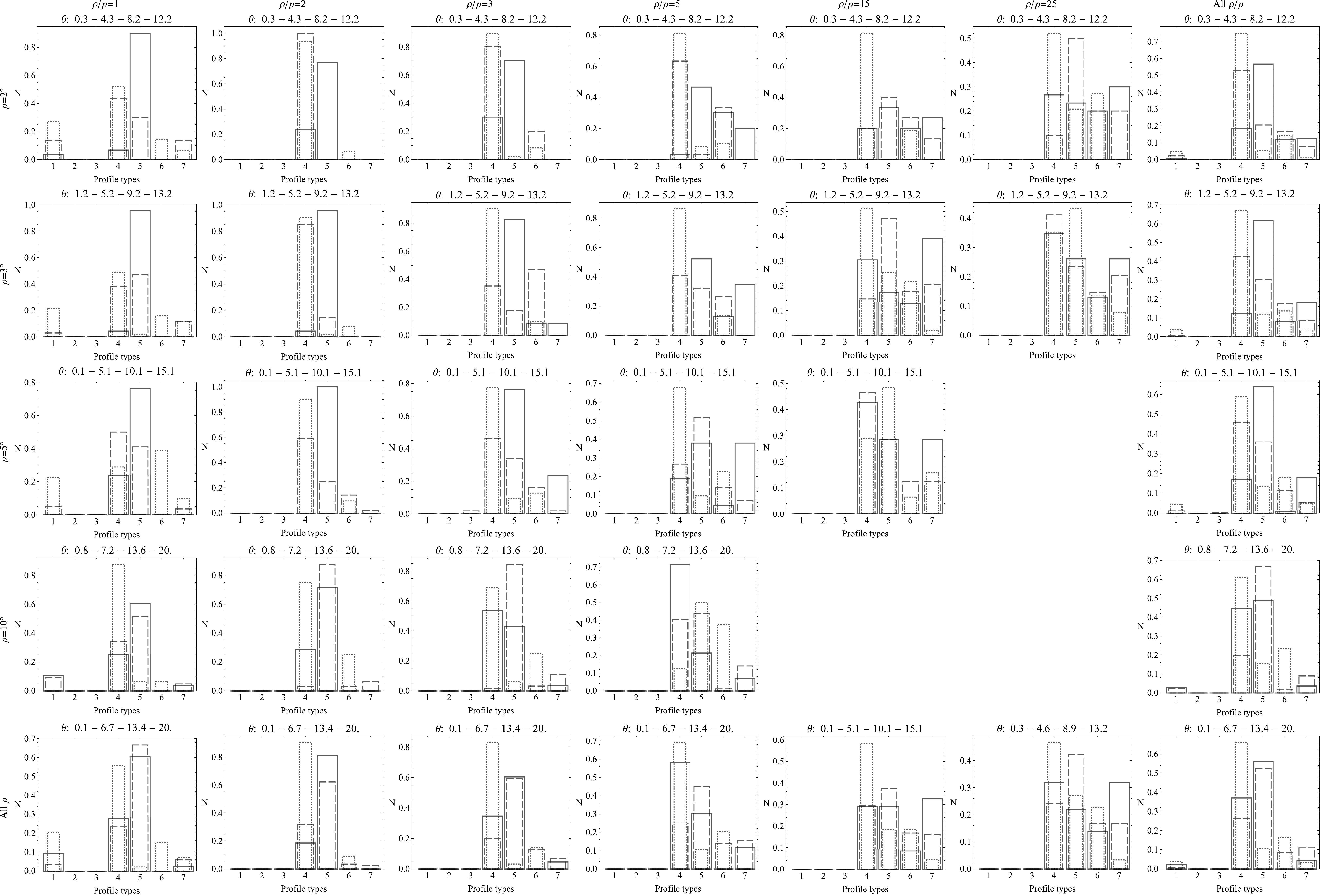, scale=0.35}}
\caption{Shapes of EV distributions depending on the angle of the velocity vector ($\theta$) and the jet component axis ($\theta_\rho$) to the line of sight for $\psi^\prime=55^\circ$. Solid, dashed, and dotted lines are associated with intervals of high, medium, and small values, respectively. The intervals of $\theta$ are indicated at the top of each plot.}
\label{fig:fig9}
\end{sidewaysfigure}

\begin{sidewaysfigure}
\centerline{\epsfig{file=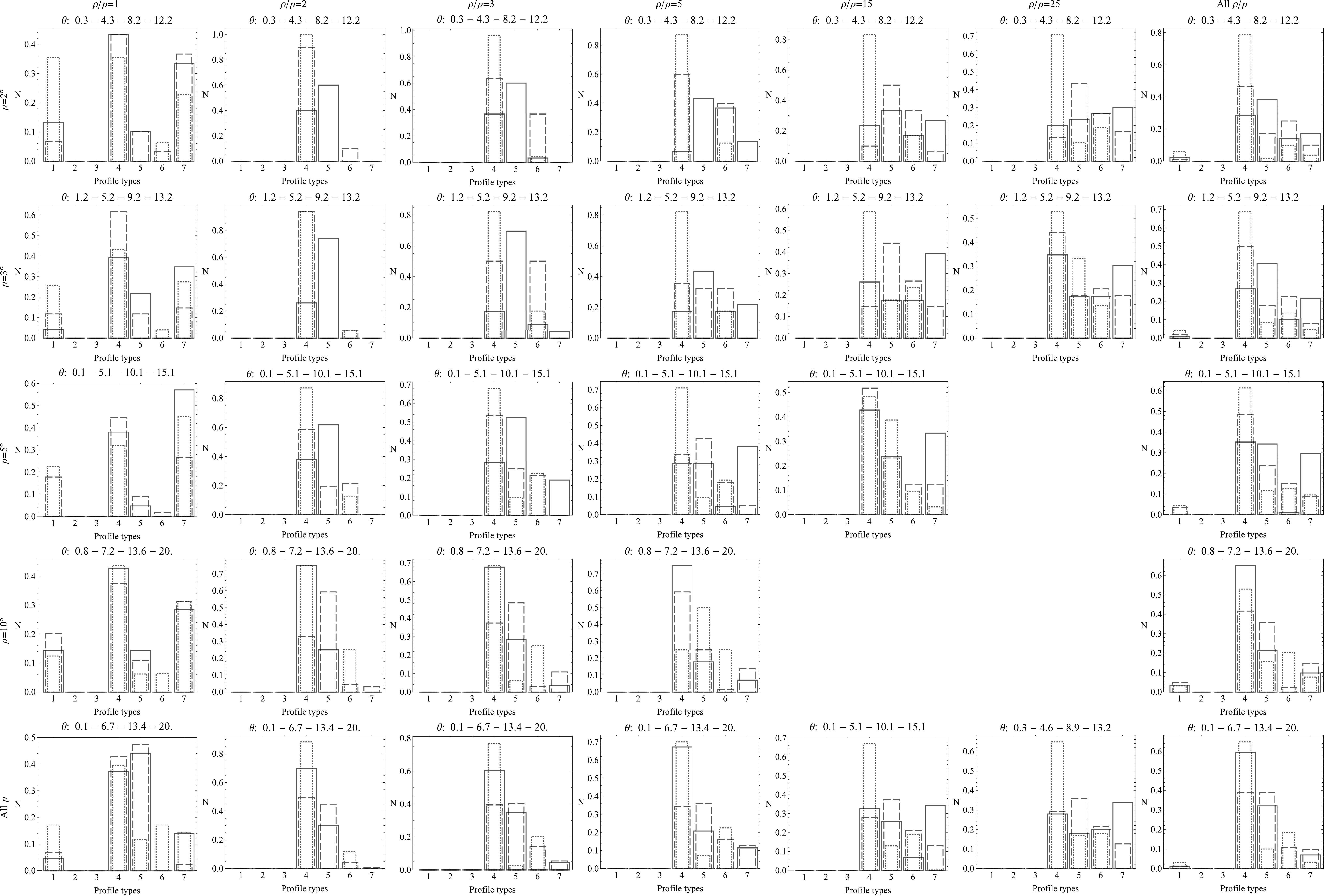, scale=0.35}}
\caption{Shapes of EV distributions depending on the angle of the velocity vector ($\theta$) and the jet component axis ($\theta_\rho$) to the line of sight for $\psi^\prime=65^\circ$. Solid, dashed, and dotted lines are associated with intervals of high, medium, and small values, respectively. The intervals of $\theta$ are indicated at the top of each plot.}
\label{fig:fig10}
\end{sidewaysfigure}

\begin{sidewaysfigure}
\centerline{\epsfig{file=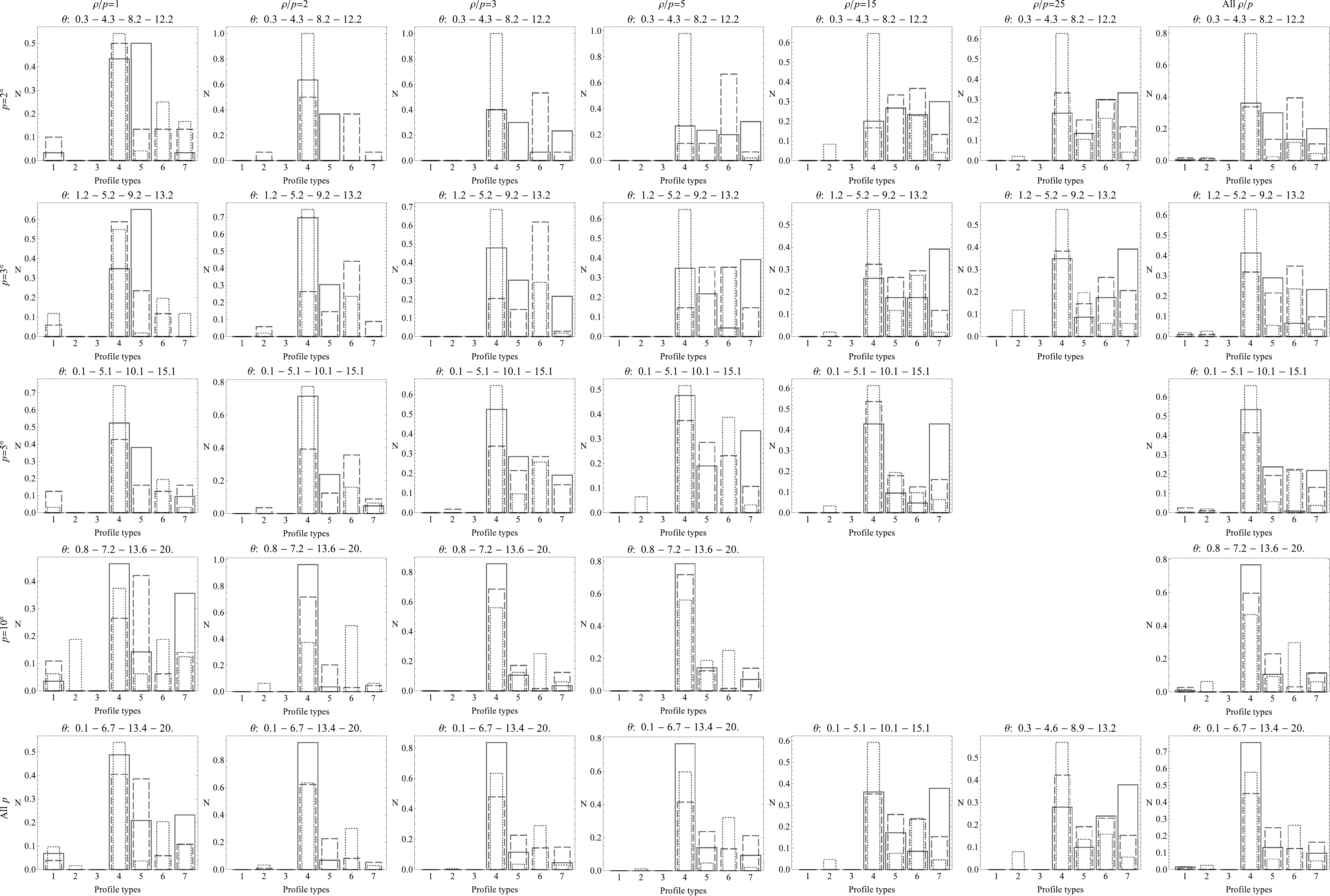, scale=0.35}}
\caption{Shapes of EV distributions depending on the angle of the velocity vector ($\theta$) and the jet component axis ($\theta_\rho$) to the line of sight for $\psi^\prime=75^\circ$. Solid, dashed, and dotted lines are associated with intervals of high, medium, and small values, respectively. The intervals of $\theta$ are indicated at the top of each plot.}
\label{fig:fig11}
\end{sidewaysfigure}

\begin{sidewaysfigure}
\centerline{\epsfig{file=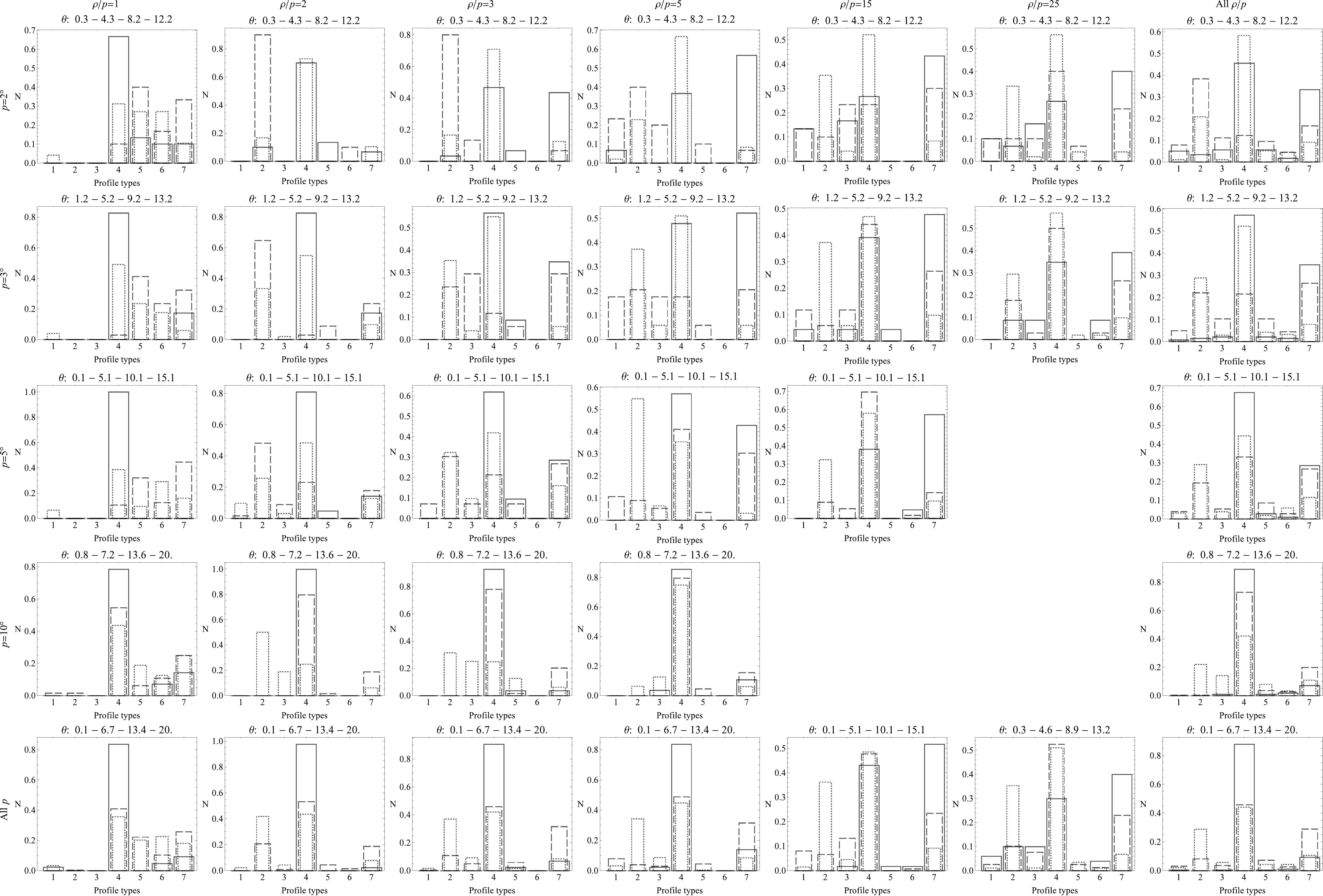, scale=0.35}}
\caption{Shapes of EV distributions depending on the angle of the velocity vector ($\theta$) and the jet component axis ($\theta_\rho$) to the line of sight for $\psi^\prime=90^\circ$. Solid, dashed, and dotted lines are associated with intervals of high, medium, and small values, respectively. The intervals of $\theta$ are indicated at the top of each plot.}
\label{fig:fig12}
\end{sidewaysfigure}

\begin{sidewaysfigure}
\centerline{\epsfig{file=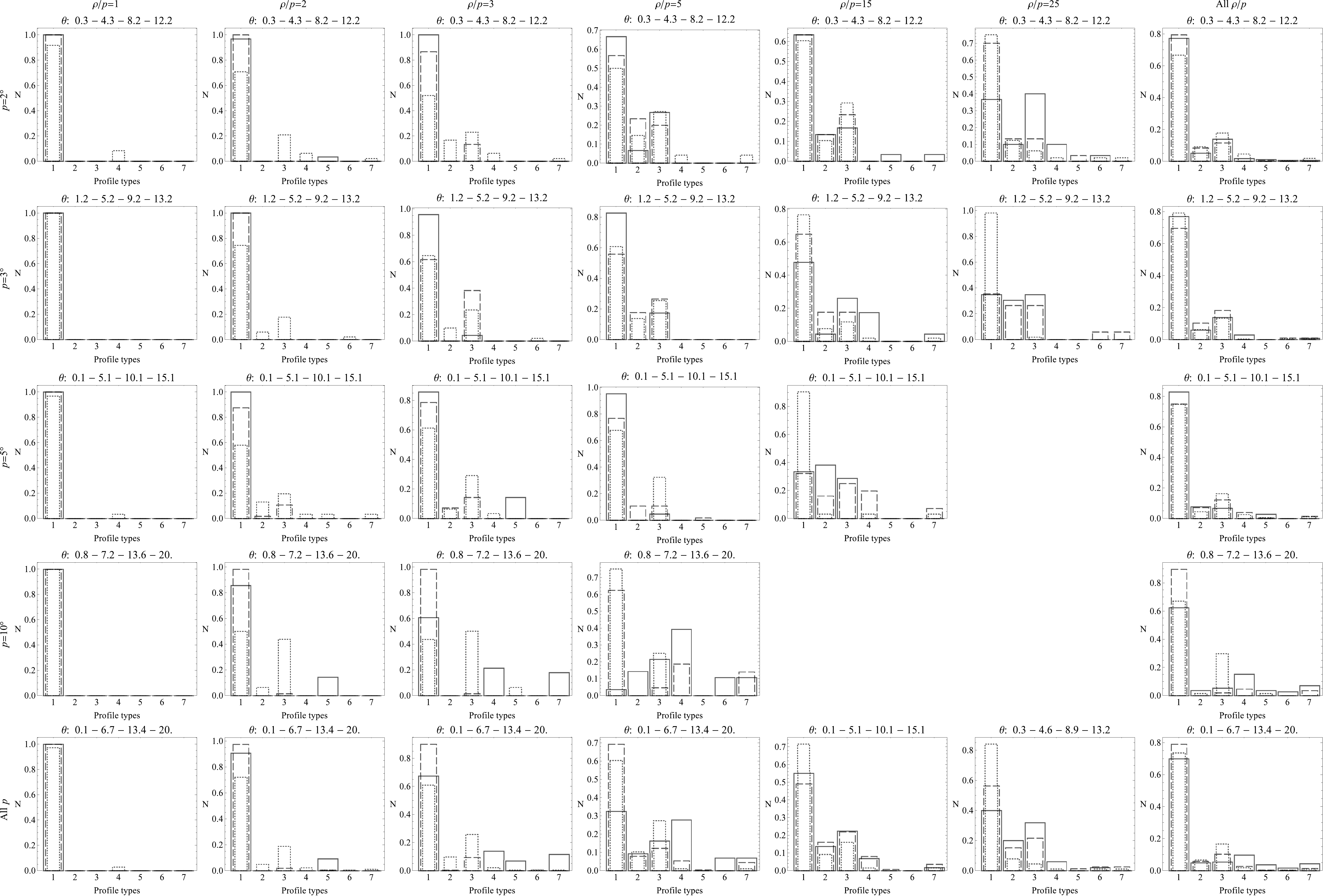, scale=0.35}}
\caption{
The shapes of the EV distributions depend on the angle of the velocity vector ($\theta$) and jet component axis ($\theta_\rho$) to the line of sight for $R_t=0.25$. Solid, dashed, and dotted lines are associated with intervals of high, medium and small values of $\theta$, respectively. The intervals of $\theta$ are indicated at the top of each plot.
}
\label{fig:fig13}
\end{sidewaysfigure}

\clearpage
\begin{sidewaysfigure}
\centerline{\epsfig{file=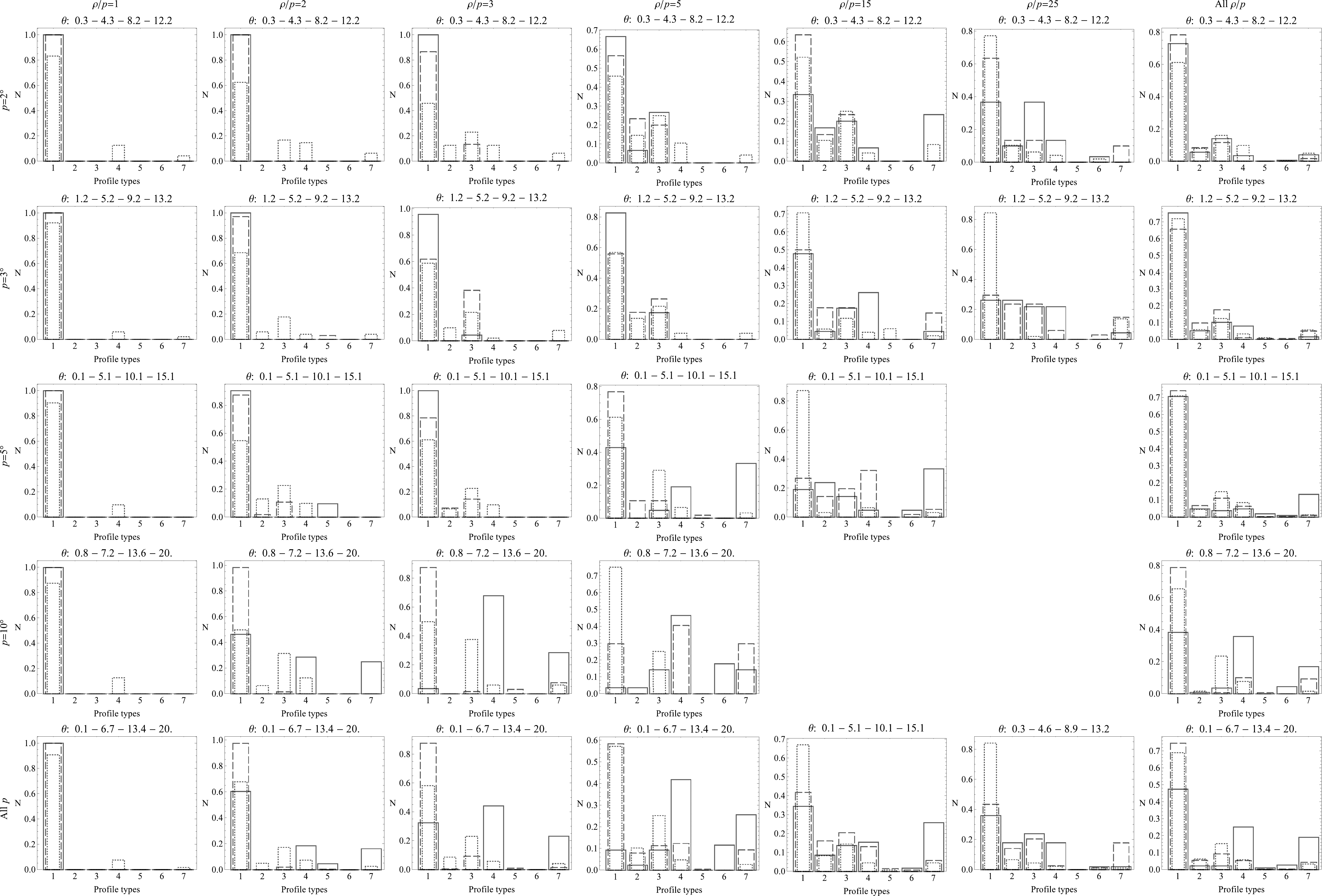, scale=0.35}}
\caption{
The shapes of the EV distributions depend on the angle of the velocity vector ($\theta$) and jet component axis ($\theta_\rho$) to the line of sight for $R_t=0.33$. Solid, dashed, and dotted lines are associated with intervals of high, medium and small values of $\theta$, respectively. The intervals of $\theta$ are indicated at the top of each plot.}
\label{fig:fig14}
\end{sidewaysfigure}

\begin{sidewaysfigure}
\centerline{\epsfig{file=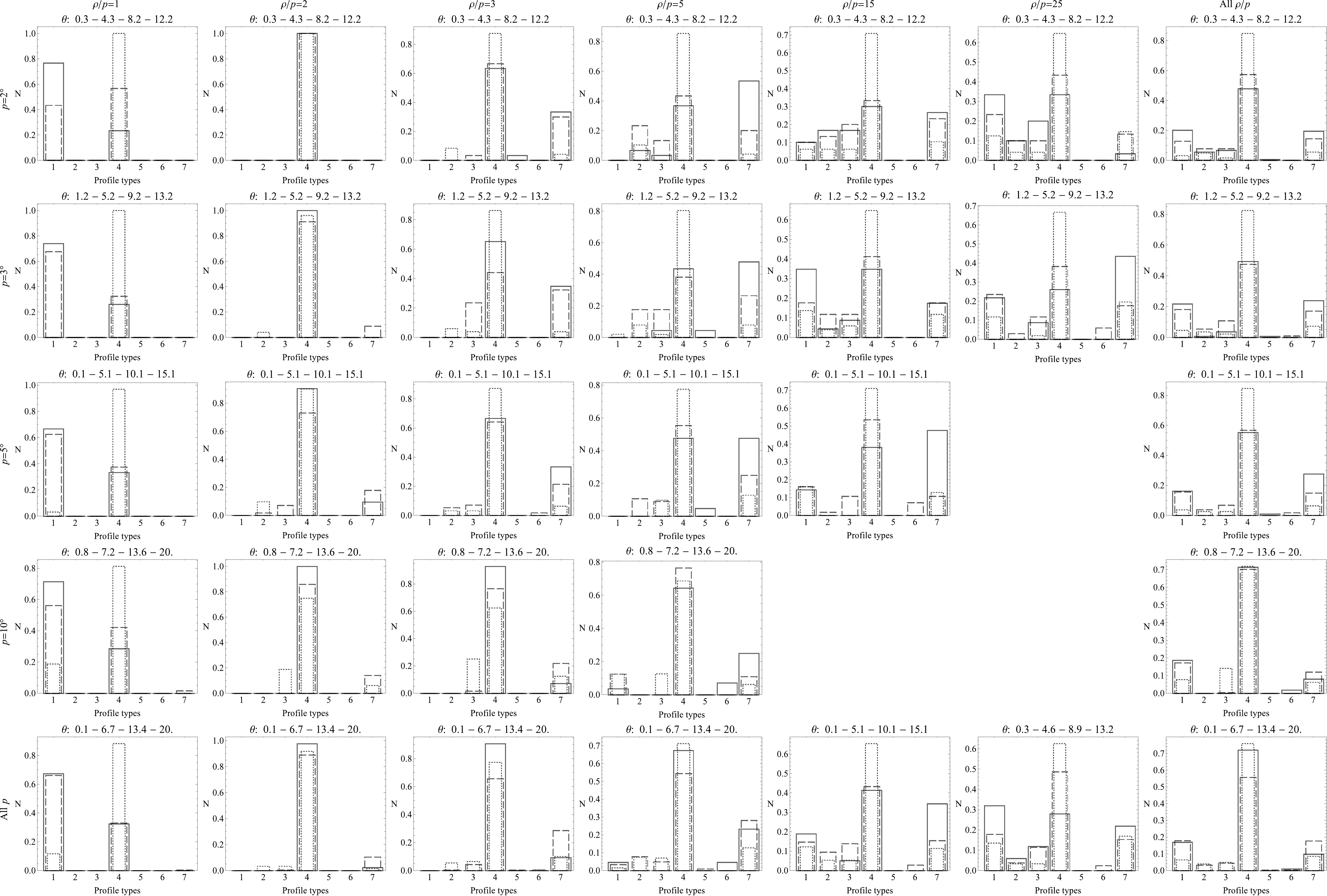, scale=0.35}}
\caption{
The shapes of the EV distributions depend on the angle of the velocity vector ($\theta$) and jet component axis ($\theta_\rho$) to the line of sight for $R_t=0.5$. Solid, dashed, and dotted lines are associated with intervals of high, medium and small values of $\theta$, respectively. The intervals of $\theta$ are indicated at the top of each plot.}
\label{fig:fig15}
\end{sidewaysfigure}

\begin{sidewaysfigure}
\centerline{\epsfig{file=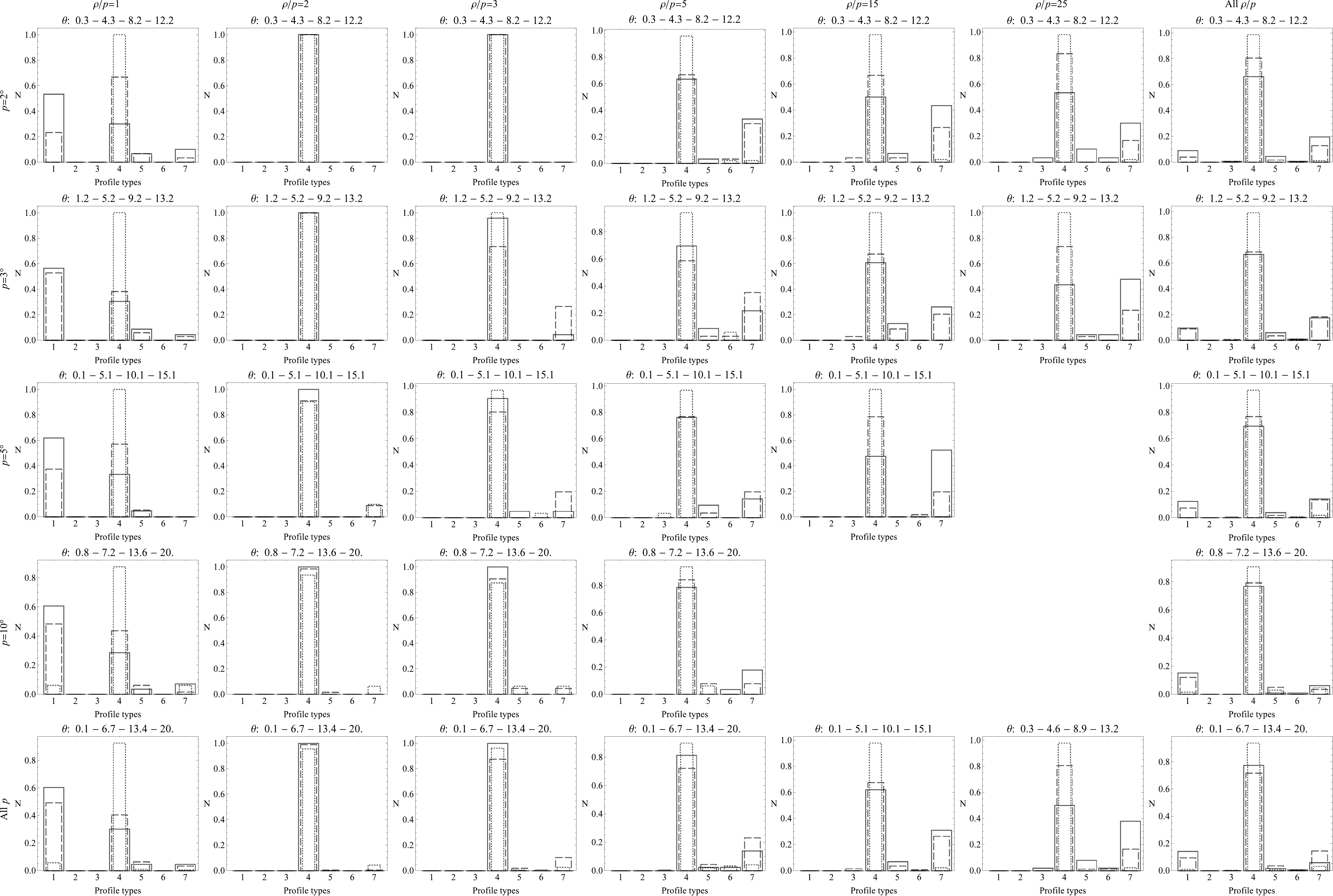, scale=0.35}}
\caption{
The shapes of the EV distributions depend on the angle of the velocity vector ($\theta$) and jet component axis ($\theta_\rho$) to the line of sight for $R_t=0.7$. Solid, dashed, and dotted lines are associated with intervals of high, medium and small values of $\theta$, respectively. The intervals of $\theta$ are indicated at the top of each plot.}
\label{fig:fig16}
\end{sidewaysfigure}

\begin{sidewaysfigure}
\centerline{\epsfig{file=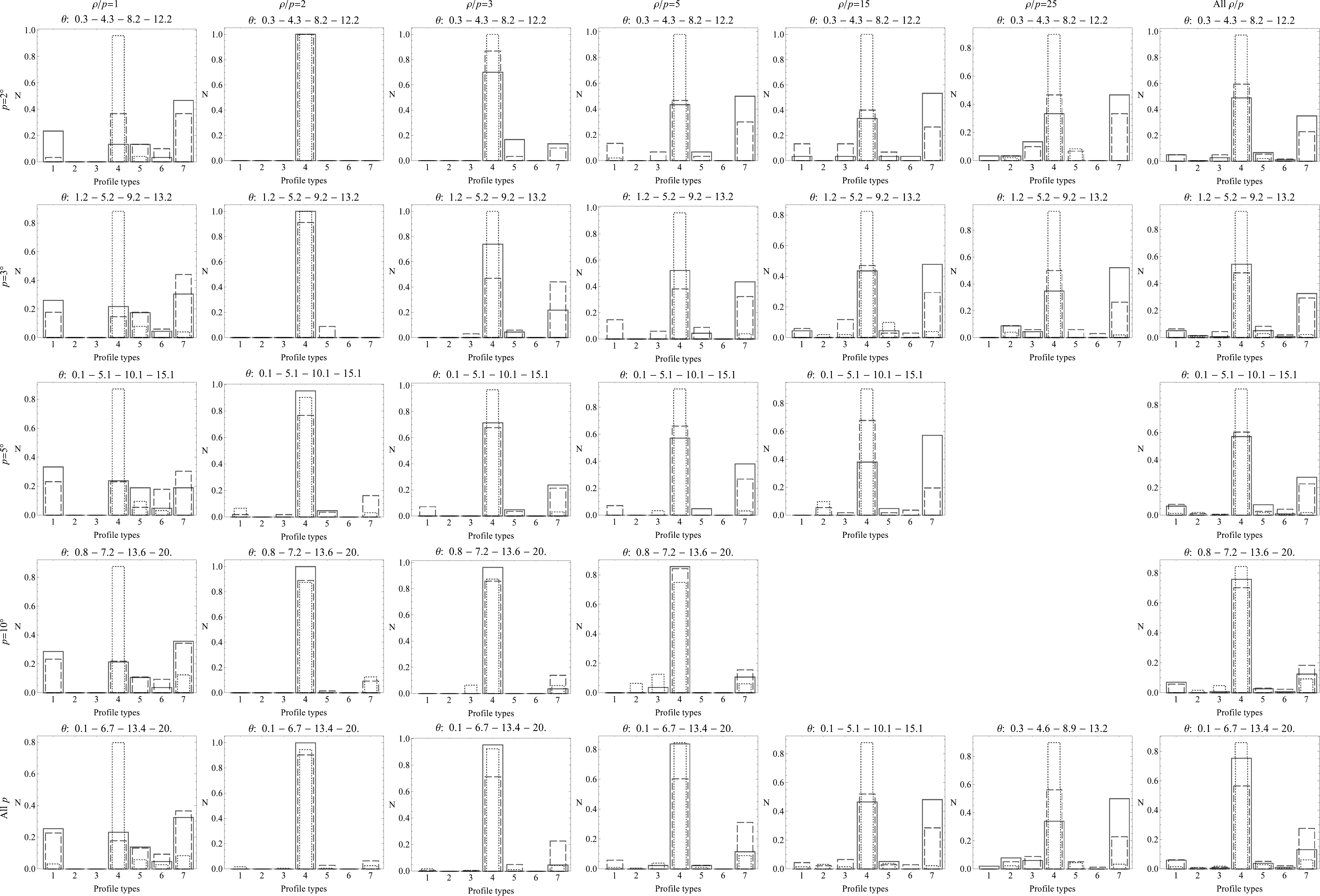, scale=0.35}}
\caption{
The shapes of the EV distributions depend on the angle of the velocity vector ($\theta$) and jet component axis ($\theta_\rho$) to the line of sight for $R_t=0.9$. Solid, dashed, and dotted lines are associated with intervals of high, medium and small values of $\theta$, respectively. The intervals of $\theta$ are indicated at the top of each plot.}
\label{fig:fig17}
\end{sidewaysfigure}

\begin{figure}[hb]
\includegraphics[scale=0.8]{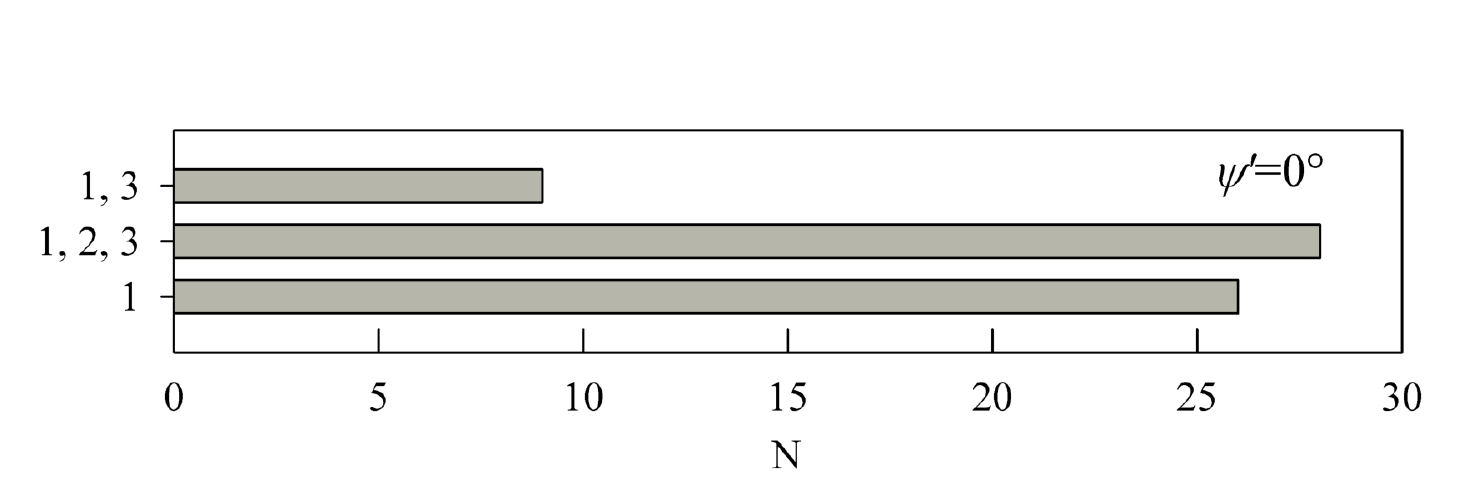} 
\caption{Combination of EV distribution shapes in each model jet for $\psi^\prime=0^\circ$. }  \label{fig:fig18}
\end{figure}

\begin{figure}
\includegraphics[scale=0.8]{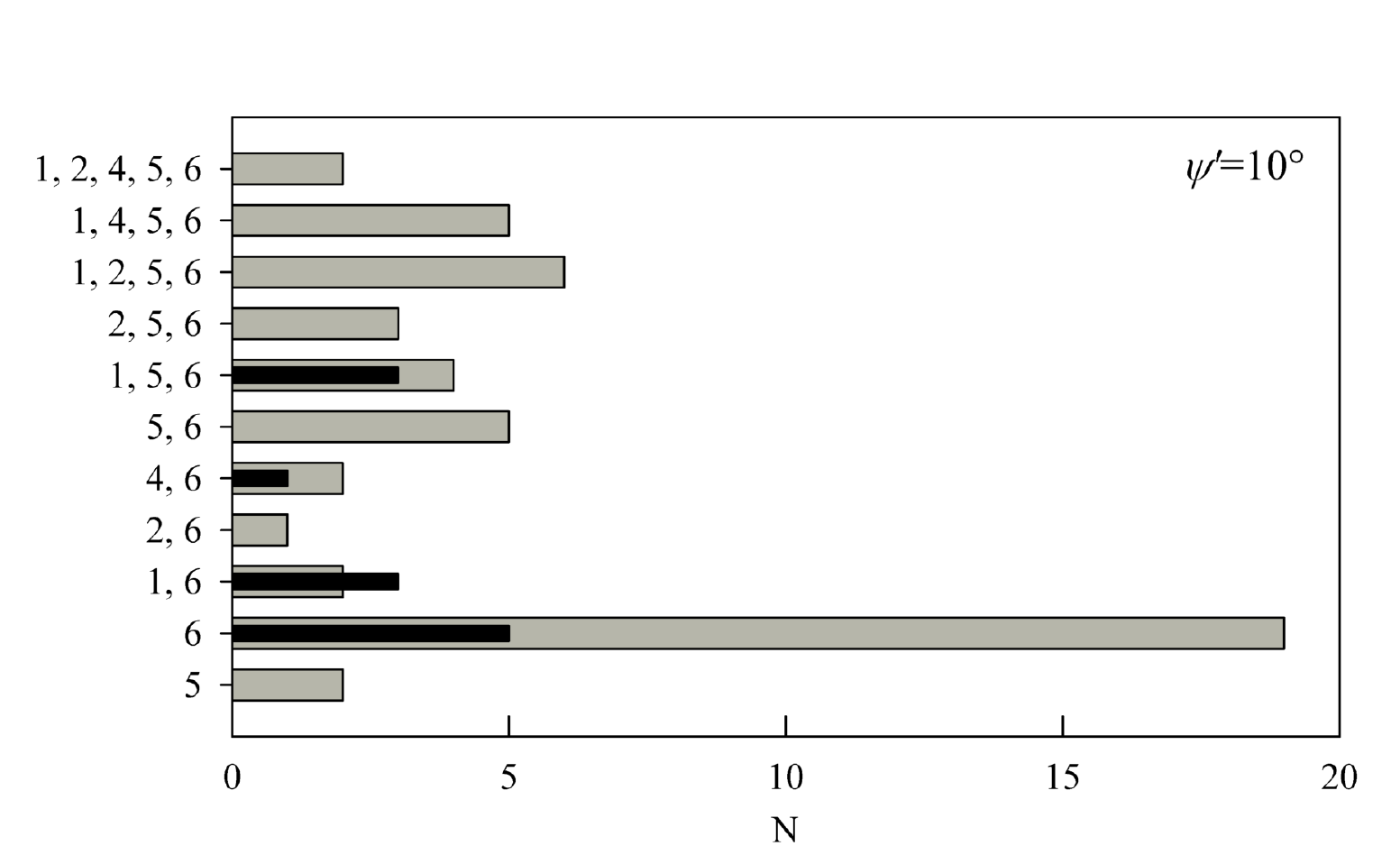} 
\caption{Combination of EV distribution shapes in each model jet for $\psi^\prime=10^\circ$. Those that additionally contain type~7 are marked in black.}        \label{fig:fig19}
\end{figure}

\begin{figure}
\includegraphics[scale=0.8]{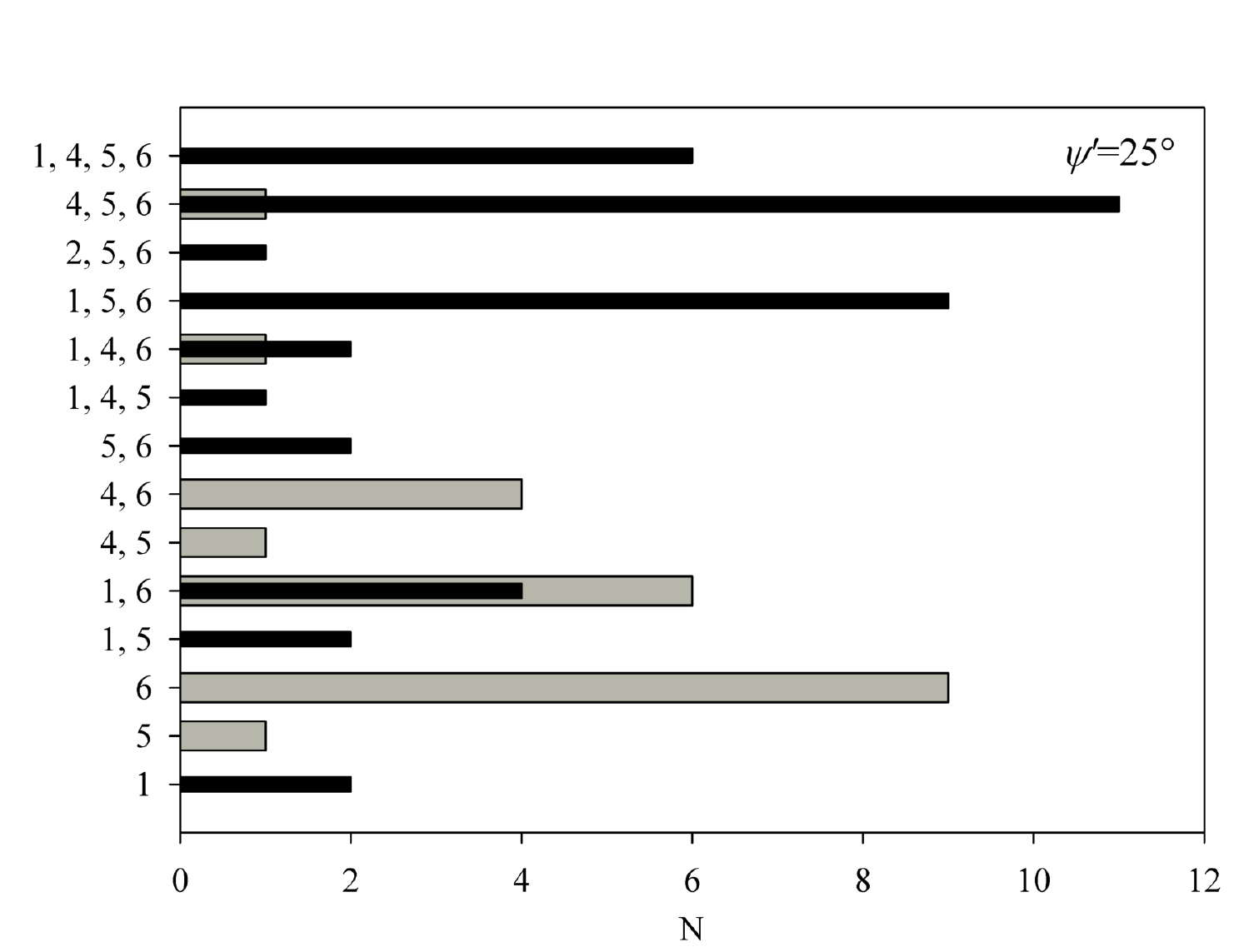} 
\caption{Combination of EV distribution shapes in each model jet for $\psi^\prime=25^\circ$. Those that additionally contain type~7 are marked in black.}     \label{fig:fig20}
\end{figure}

\begin{figure}
\includegraphics[scale=0.8]{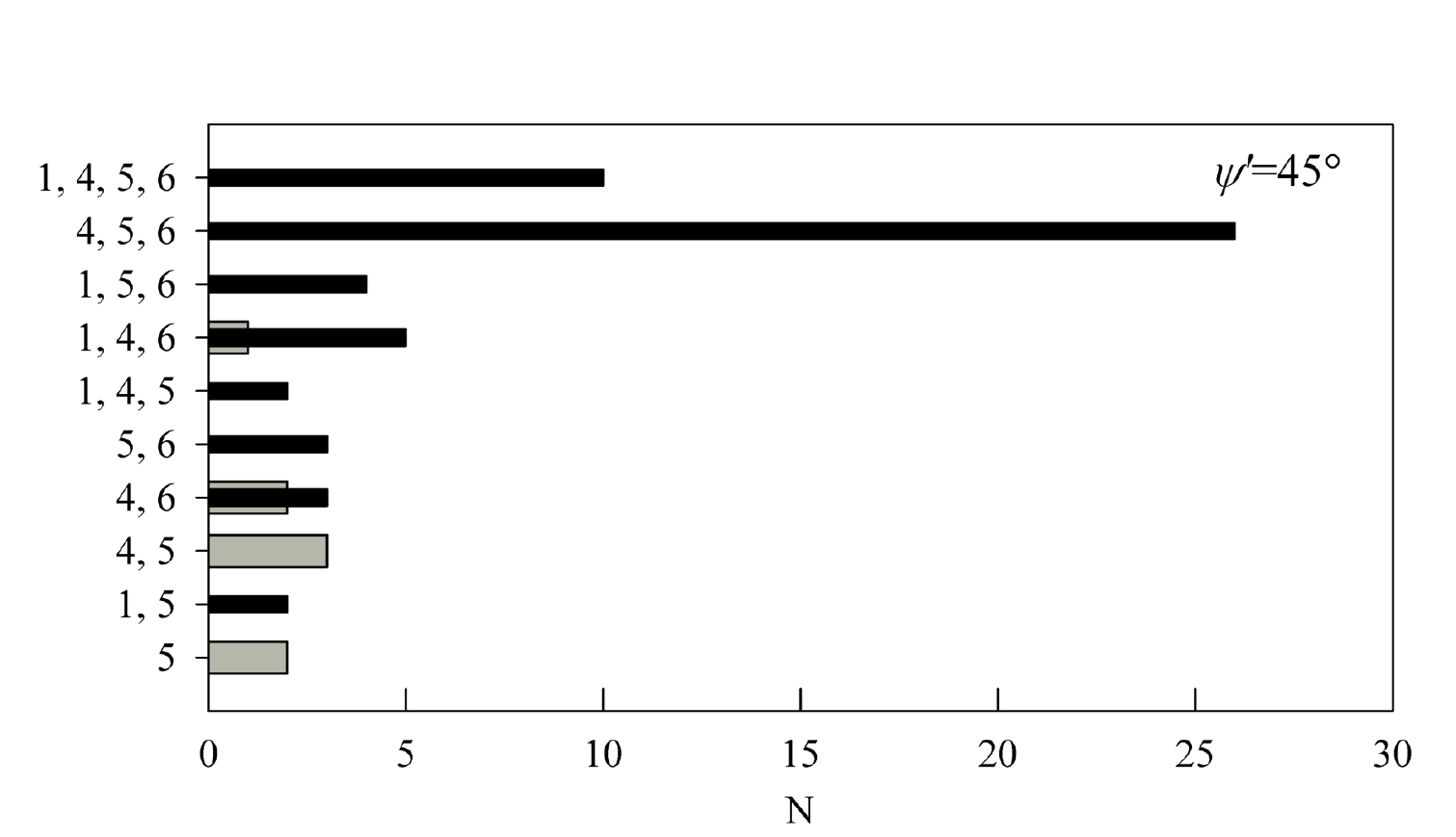} 
\caption{Combination of EV distribution shapes in each model jet for $\psi^\prime=45^\circ$. Those that additionally contain type~7 are marked in black.}        \label{fig:fig21}
\end{figure}

\begin{figure}
\includegraphics[scale=0.8]{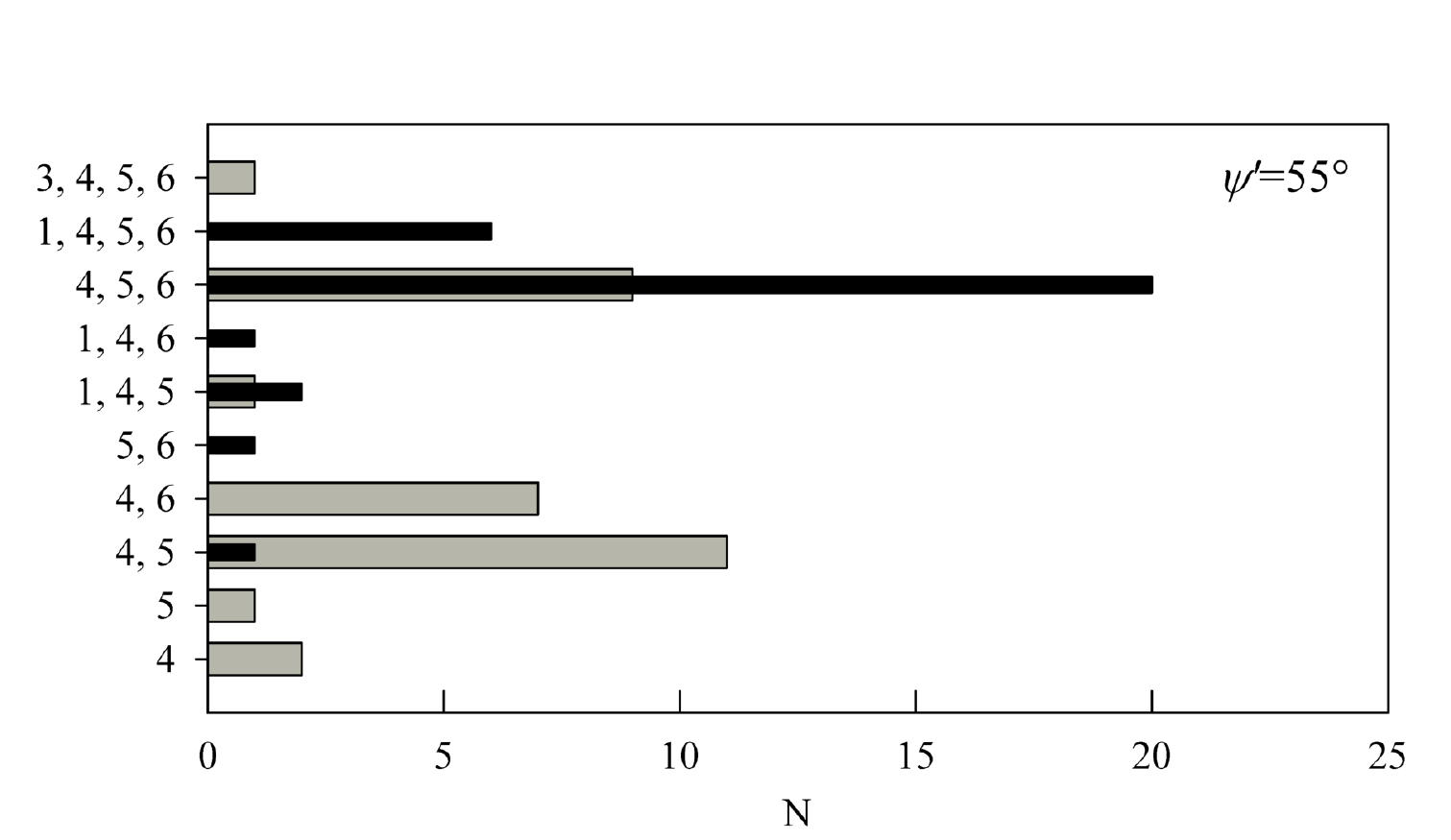} 
\caption{Combination of EV distribution shapes in each model jet for $\psi^\prime=55^\circ$. Those that additionally contain type~7 are marked in black.}     \label{fig:fig22}
\end{figure}

\begin{figure} 
\includegraphics[scale=0.8]{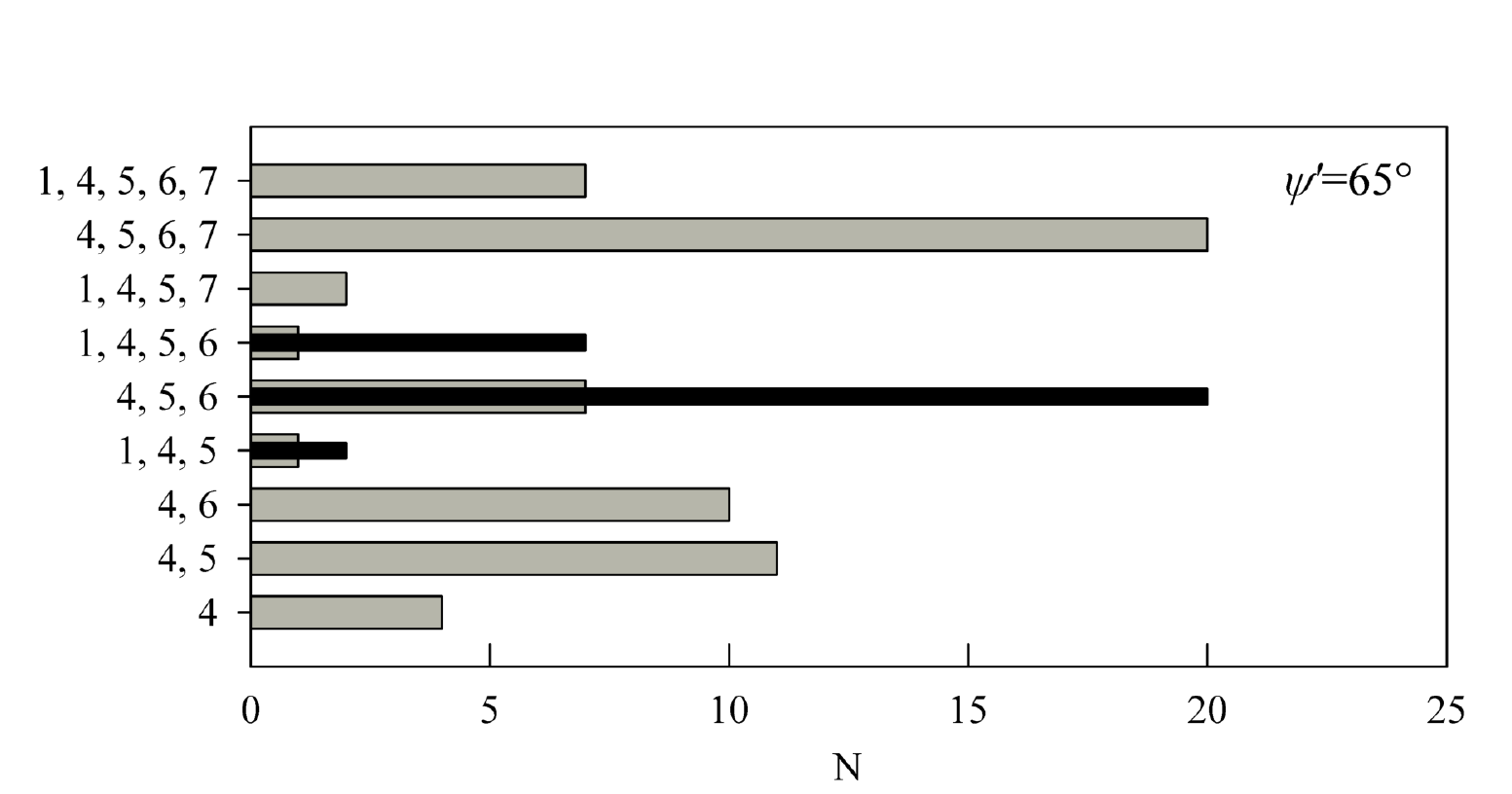} 
\caption{Combination of EV distribution shapes in each model jet for $\psi^\prime=65^\circ$. Those that additionally contain type~7 are marked in black.}               \label{fig:fig23}
\end{figure}

\begin{figure}
\includegraphics[scale=0.8]{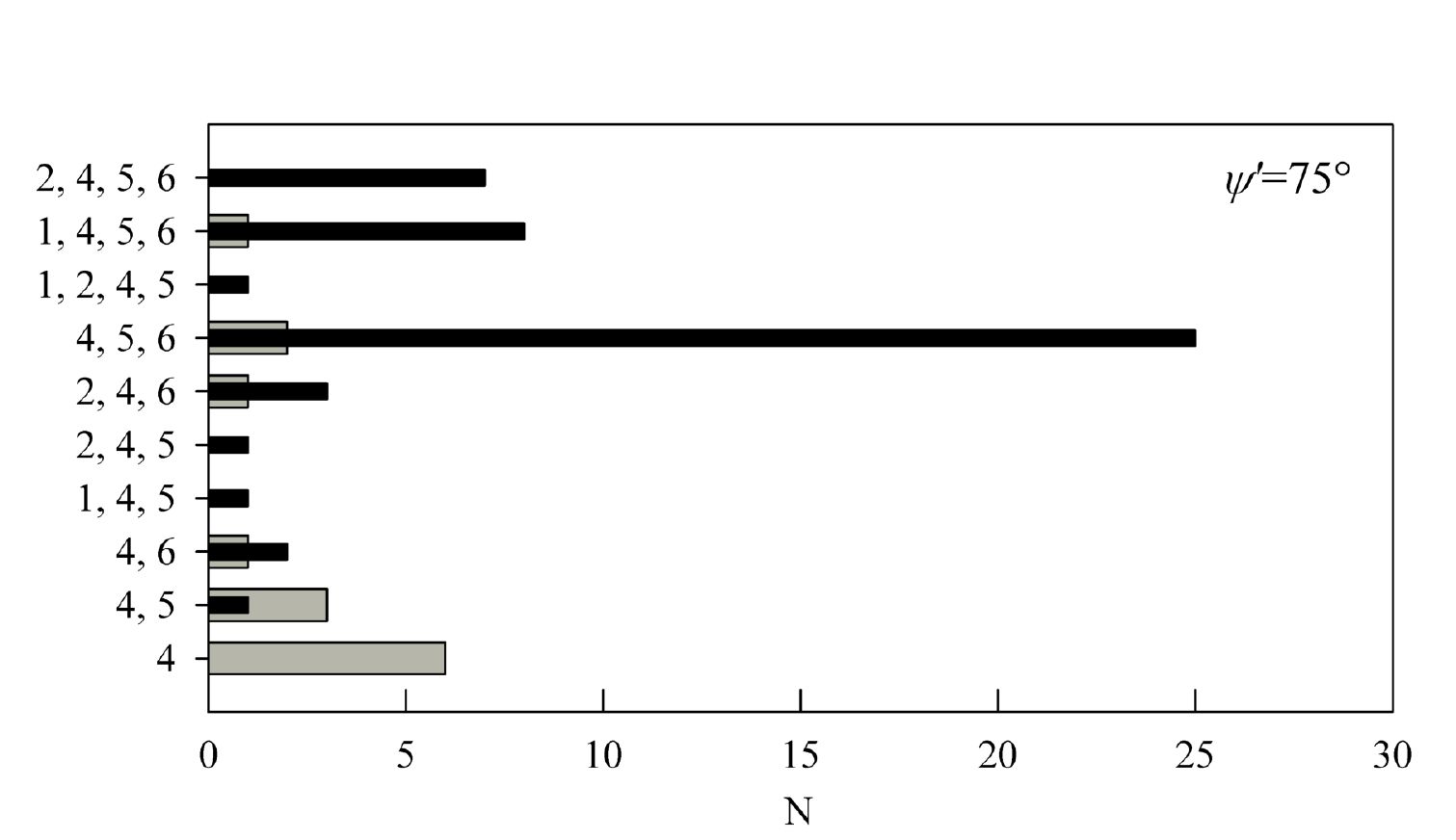} 
\caption{Combination of EV distribution shapes in each model jet for $\psi^\prime=75^\circ$. Those that additionally contain type~7 are marked in black.}           \label{fig:fig24}
\end{figure}

\begin{figure}
\includegraphics[scale=0.8]{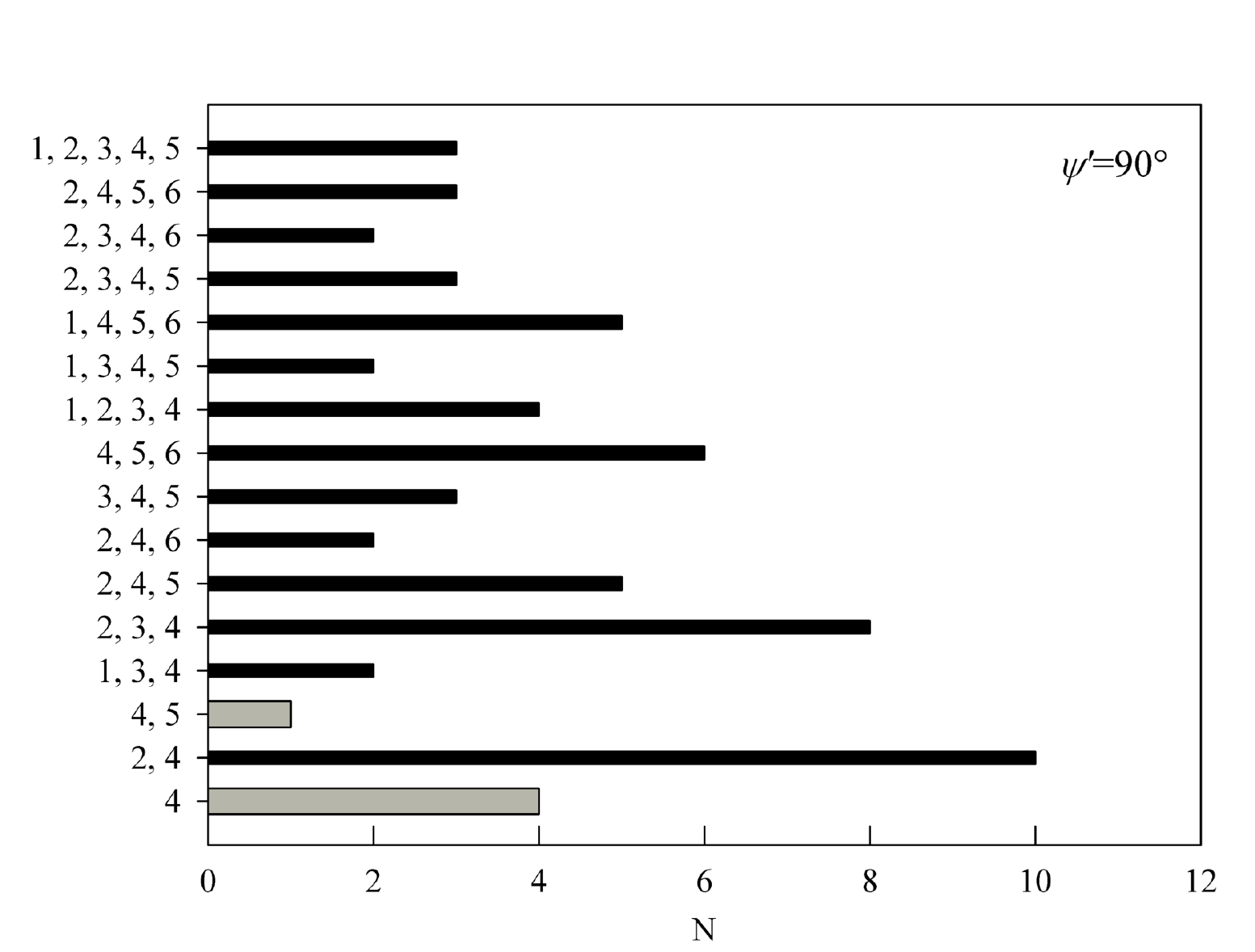} 
\caption{Combination of EV distribution shapes in each model jet for $\psi^\prime=90^\circ$. Those that additionally contain type~7 are marked in black.}        \label{fig:fig25}
\end{figure}

\begin{figure}
\includegraphics[scale=0.8]{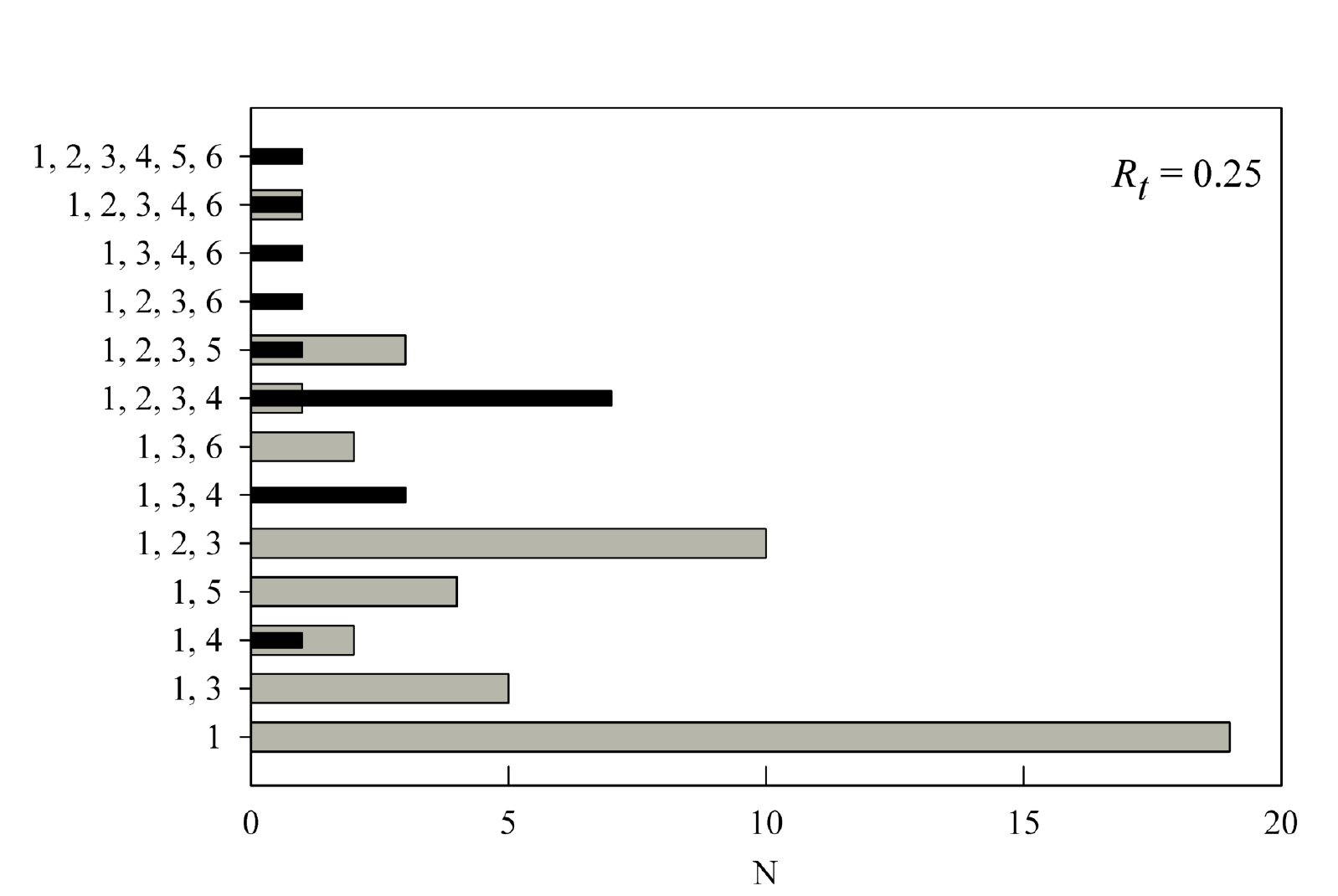} 
\caption{Combination of EV distribution shapes in each model jet for$R_t=0.25$. Those that additionally contain type~7 are marked in black.}        \label{fig:fig26}
\end{figure}

\begin{figure}
\includegraphics[scale=0.8]{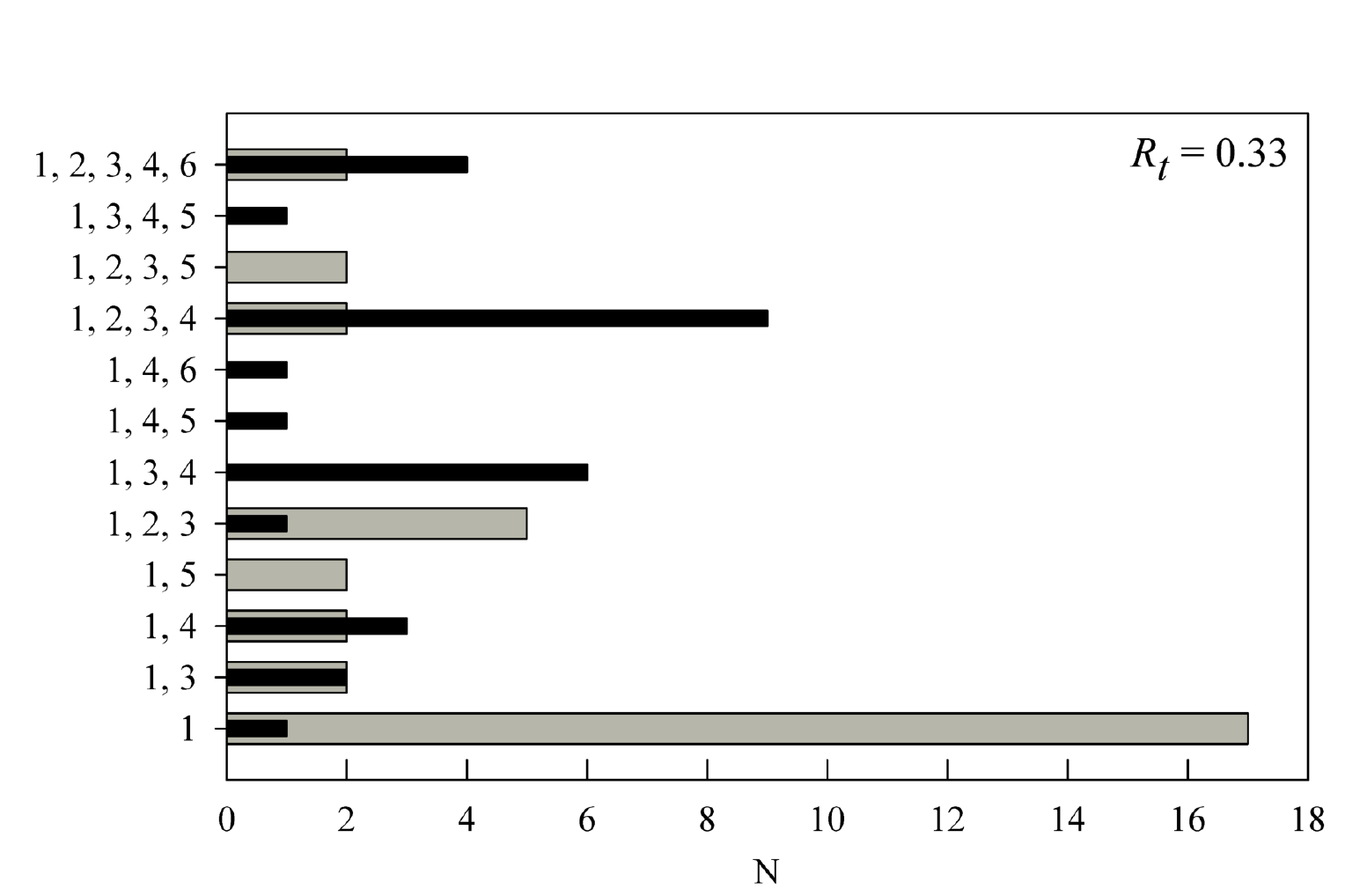} 
\caption{Combination of EV distribution shapes in each model jet for$R_t=0.33$. Those that additionally contain type~7 are marked in black.}          \label{fig:fig27}
\end{figure}

\begin{figure}
\includegraphics[scale=0.8]{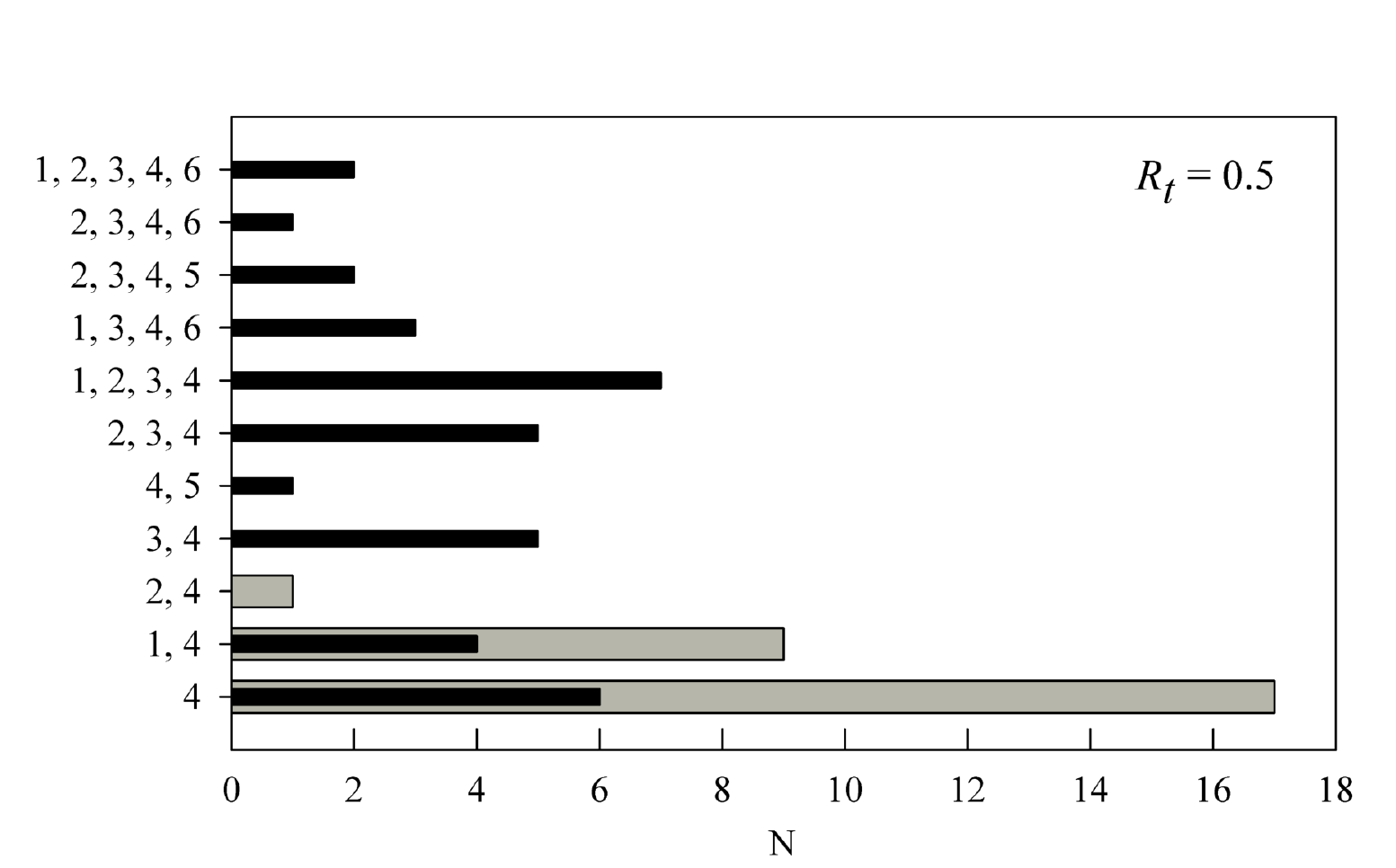} 
\caption{Combination of EV distribution shapes in each model jet for$R_t=0.5$. Those that additionally contain type~7 are marked in black.}         \label{fig:fig28}
\end{figure}

\begin{figure}
\includegraphics[scale=0.8]{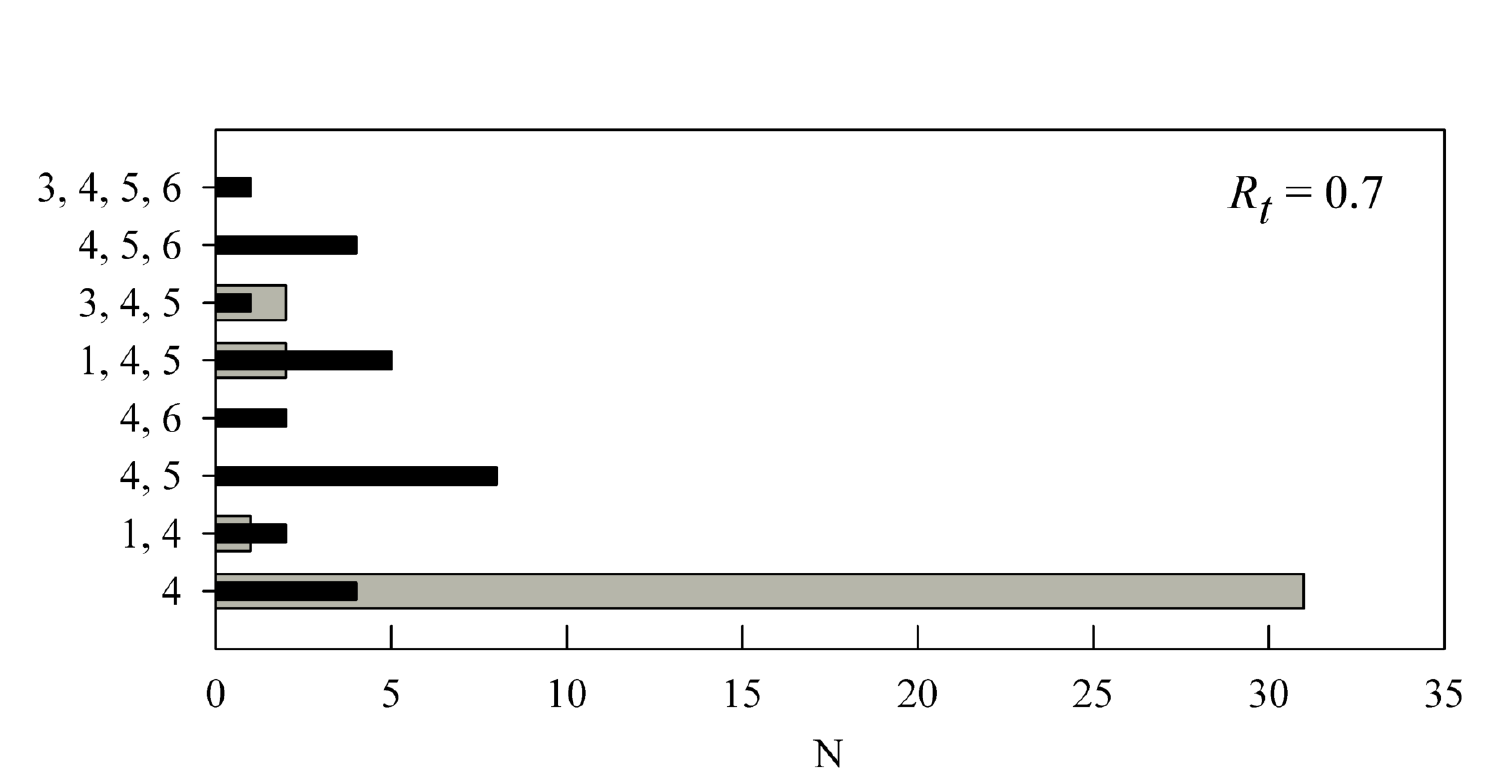} 
\caption{Combination of EV distribution shapes in each model jet for$R_t=0.7$. Those that additionally contain type~7 are marked in black.}         \label{fig:fig29}
\end{figure}

\begin{figure}
\includegraphics[scale=0.8]{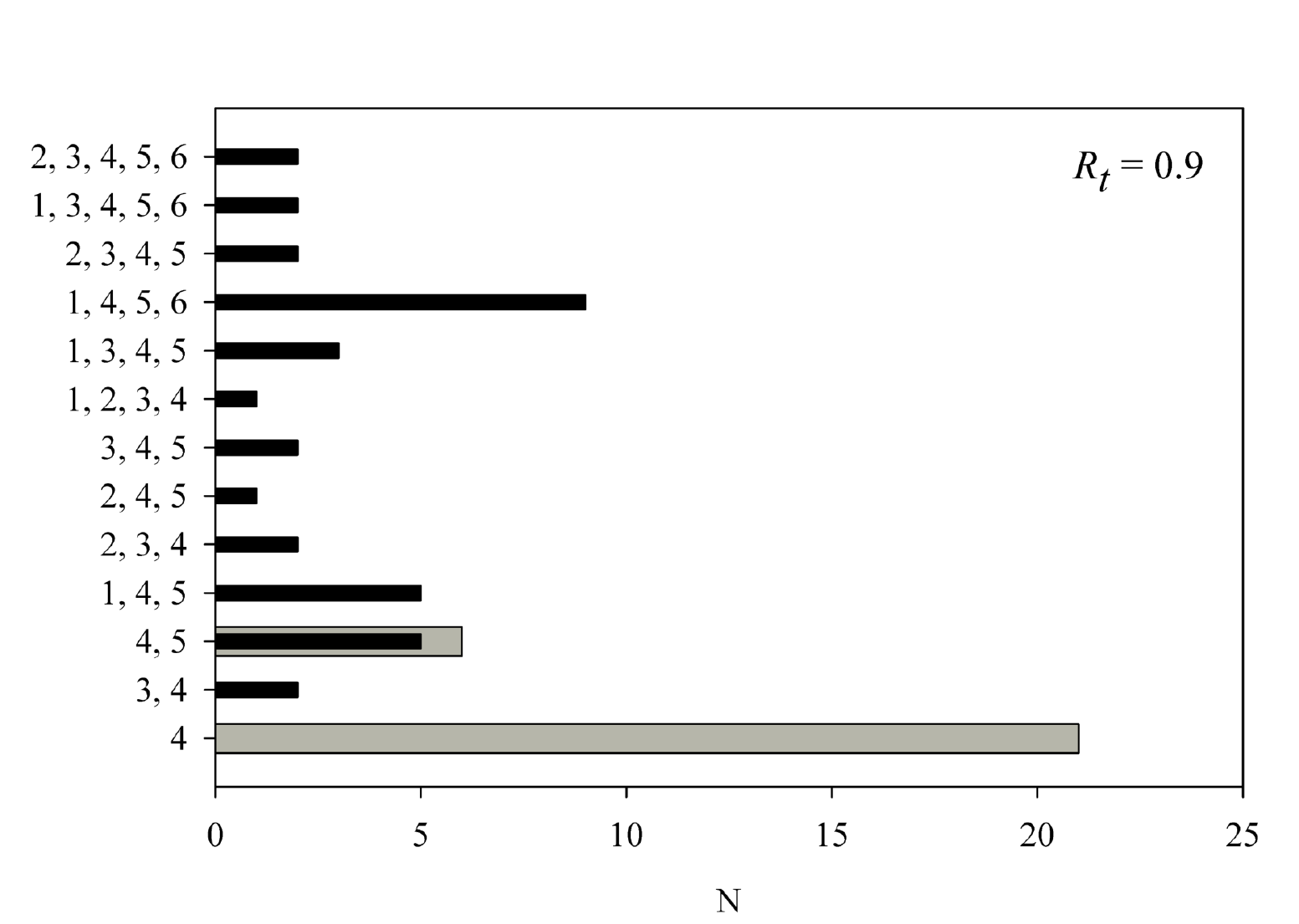} 
\caption{Combination of EV distribution shapes in each model jet for$R_t=0.9$. Those that additionally contain type~7 are marked in black.}         \label{fig:fig30}
\end{figure}

\end{document}